# Eulerian−Lagrangian modelling of rotating detonative combustion in partially pre-vaporized *n*-heptane sprays with hydrogen addition

Qingyang Meng[a,b], Majie Zhao[b], Hongtao Zheng[a], Huangwei Zhang[b,*]

[a] *College of Power and Energy Engineering, Harbin Engineering University, Harbin 150001, China*

[b] *Department of Mechanical Engineering, National University of Singapore, 9 Engineering Drive 1, Singapore 117576, Republic of Singapore*

**Abstract**

Rotating detonation combustion (RDC) fuelled with partially pre-vaporized *n*-heptane sprays and gaseous hydrogen is studied with an Eulerian−Lagrangian method. Our focus is the effects of pre-vaporized *n*-heptane equivalence ratios and droplet diameters on detonation wave propagation and droplet dynamics in two-phase RDC. The results show that when the droplets are small, they are fully vaporized by the detonation wave. However, when the droplet diameter is relatively large and/or the detonation wave number is bifurcated, liquid droplets are observable beyond the refill zone. Moreover, the detonation speed is considerably influenced by the droplet pre-vaporization and diameter. The velocity deficits vary between 5% and 30%. Over 70% *n*-heptane is detonated in the simulated cases, and there exists a critical droplet diameter (about 20 μm), around which the detonated fuel fraction is minimal. Four droplet trajectories in RDC are identified, which are differentiated by various evaporation times, residence times and interactions between droplets and the basic RDC flow structures. Inside the refill zone, three droplet categories are qualitatively identified. Droplets injected at the right end of the refill zone directly interact with the deflagration surface and meanwhile have relatively long residence time. However, droplets injected closer to the travelling detonation front have insufficient time to be heated and vaporized. Our results also demonstrate that when pre-vaporization level is low and initial droplet diameter is large, the liquid fuel droplets may disperse towards the combustor exit. Furthermore, the droplet dispersion height decreases with liquid fuel pre-vaporization, while increases with droplet diameter.

**Keywords:** Rotating detonation combustion, *n*-heptane, hydrogen, fuel spray, pre-vaporization, Eulerian−Lagrangian modelling

[*]Corresponding author. Tel.: +65 6516 2557; Fax: +65 6779 1459.
*E-mail address*: huangwei.zhang@nus.edu.sg.

# 1. Introduction

Propulsive devices based on detonation combustion have attracted increased interests from the scientific and engineering communities in recent years, due to their higher thermodynamic efficiencies than those of deflagration-dominated engines [1, 2]. Among the different detonation engines, Rotating Detonation Engines (RDE) have numerous advantages, including simple ignition process, compact engine geometry and steady thrust output. Extensive studies have been conducted on RDE's by means of experiments [3-5], numerical simulations [6-13] and theoretical analysis [14-17].

Most of the existing RDE studies use gaseous fuels (e.g. hydrogen or ethylene) [1, 2]. However, for practical applications, liquid fuels are more desirable, due to their small storage space, high energy density and extensive supply source. Early interests in liquid fuelled RDE's date back to 1960s−70s [16], which was mainly motivated by rocket propulsion technology development for space exploration in that era. In recent years, they have been revived, to develop novel pressure-gain combustion technology, and a series of laboratory-scale RDE experiments have been successfully realized with various liquid fuels. For instance, Bykovskii et al. [18-20] achieved two-phase Rotating Detonation Combustion (RDC) by spraying liquid kerosene. In their tests with an annular cylindrical combustor (diameter 306 mm), the oxygen-enriched air should be used as the oxidant to initiate the detonation [18]. More recently, hydrogen or syngas is added in their experiments with a larger combustor (diameter 503 mm) using standard air and kerosene sprays [19, 20]. In this combustor, Rotating Detonation Waves (RDW's) cannot be initiated without gaseous hydrogen or syngas. They systematically discussed the RDW propagation characteristics and propulsive performance of liquid kerosene fuelled RDE's [19, 20].

In addition, Kindracki [21] investigated kerosene atomization characteristics in cold nitrogen flows by changing nitrogen velocity and fuel injection pattern of a model detonation chamber. They found that most of the droplets in their experiments have diameters of 20−40 µm, which can quickly evaporate and form a combustible mixture in the chamber. Kindracki [22] used liquid kerosene and gaseous air to study initiation and propagation of RDW's. In his work, continuous propagation of detonation wave was successfully achieved for a mixture of liquid kerosene and air and a small addition of hydrogen. Velocity deficits of 20−25% are observed for his studied heterogeneous mixtures.

Li et al. [23] investigated the RDW behaviours in both premixed and non-premixed injection schemes by using vaporized Jet A-1 and pre-heated air. RDW's stabilize with both injection methods. They also tested the RDC at cold starting, i.e. without air pre-heating, and it turns out that the detonation initiation is not successful, characterized by non-periodic pressure signals within the operating period. Furthermore, based on a rocket-type combustor, Xue et al. [24] used liquid nitrogen TetrOxide (NTO) and liquid MonomethylHydrazine (MMH) as the propellants to study the RDW propagation mode under different mass flow rates and outlet structures. This experiment is a preliminary demonstration of the feasibility of RDE's with liquid hypergolic propellants.

The foregoing experimental studies have provided significant scientific insights about the two-phase RDC. However, detailed information about detonation and flow fields (e.g. droplet evaporation and fuel vapour distributions) inside the channel are difficult to be measured. Alternatively, numerical simulation based on reacting two-phase flows is a promising tool for understanding fundamental physics and assisting practical design of liquid fuelled RDE's. For instance, based on Eulerian−Lagrangian method, Sun and Ma [9] used octane and air as the reactants to numerically investigate the effects of air total temperature and fuel inlet spacing on the two-phase RDW. They found that increasing the fuel inlet spacing results in reduction of RDW propagation speed. They also explored the upper limit of fuel inlet spacing (beyond which the RDW cannot propagate stably) and found that increasing air total temperature would increase the foregoing limit.

More recently, Hayashi et al. [25] used two-fluid method to simulate the JP-10/air RDE's with different droplet sizes (i.e. 1−10 μm) and pre-evaporation factors (i.e. 0−100%). From their work, high liquid droplet densities are found along the contact surface between the fresh and burned gas. They also observed that the non-reactive fuel pockets behind the detonation wave and highlighted the possible detonation quenching mechanisms caused by the interactions between the detonation front and fuel droplets. Furthermore, we studied the rotating detonation combustion fueled by partially pre-vaporized *n*-heptane sprays [26] and found that the detonation speed decreases with droplet sizes and tends to be constant when the diameter is greater than 30 µm. Meanwhile, the percentage of detonated *n*-heptane first decreases and then increases as the droplet diameter varies from 5 to 80 µm.

The objective of our work is to further clarify the physics of detonation propagation and droplet dynamics in liquid *n*-heptane fuelled RDE's. Specifically, we aim to work on the following three questions which are not answered in the previous work (e.g. [9, 25, 26]): (1) How do the droplet size and pre-evaporation affect the detonation propagation? (2) What are the percentages of the detonated fuels? (3) How are the droplets distributed in the RDE combustor? To this end, two-dimensional rotating detonative combustion is simulated based on a simplified RDC model and a hybrid Eulerian−Lanrangian approach. The reactants are partially pre-vaporized liquid *n*-heptane sprays with addition of stoichiometric hydrogen/air gas. The *n*-heptane droplet sizes and pre-vaporized gas equivalence ratios are varied to study their effects on the RDW characteristics and the droplet behaviours in the RDE chamber. Besides the above novel efforts, the droplet trajectories in the RDE chamber and the various timescales will also be discussed in this work.

The structure of the manuscript is organized as below. The governing equation and computational method are detailed in Section 2. The physical model and simulated cases are introduced in Section 3. The mesh sensitivity analysis and simulation results will be discussed in Sections 4 and 5, respectively. Section 6 closes the manuscript with the main conclusions.

**2. Governing equation and computational method**

*2.1 Governing equation for gas phase*

The governing equations of mass, momentum, energy, and species mass fraction, with the ideal gas equation of state, for the compressible reacting two-phase flows are solved in this work. They

respectively read

$$\frac{\partial \rho}{\partial t} + \nabla \cdot [\rho \mathbf{u}] = S_{mass}, \quad (1)$$

$$\frac{\partial (\rho \mathbf{u})}{\partial t} + \nabla \cdot [\mathbf{u}(\rho \mathbf{u})] + \nabla p + \nabla \cdot \mathbf{T} = \mathbf{S}_{mom}, \quad (2)$$

$$\frac{\partial (\rho E)}{\partial t} + \nabla \cdot [\mathbf{u}(\rho E + p)] + \nabla \cdot [\mathbf{T} \cdot \mathbf{u}] + \nabla \cdot \mathbf{j} = \dot{\omega}_T + S_{energy}, \quad (3)$$

$$\frac{\partial (\rho Y_m)}{\partial t} + \nabla \cdot [\mathbf{u}(\rho Y_m)] + \nabla \cdot \mathbf{s_m} = \dot{\omega}_m + S_{species,m}, (m = 1, \ldots M-1), \quad (4)$$

$$p = \rho RT. \quad (5)$$

In above equations, $t$ is time, $\nabla \cdot (\cdot)$ is divergence operator. $\rho$ is the density, $\mathbf{u}$ is the velocity vector, $T$ is gas temperature, $p$ is the pressure and updated from the equation of state, i.e. Eq. (5). $Y_m$ is the mass fraction of $m$-th species, $M$ is the total species number. Only $(M-1)$ equations are solved in Eq. (4) and the mass fraction of the inert species (e.g. nitrogen) can be recovered from $\sum_{m=1}^{M} Y_m = 1$. $E$ is the total energy, defined as $E = e + |\mathbf{u}|^2/2$ with $e$ being the specific internal energy. $R$ in Eq. (5) is specific gas constant and is calculated from $R = R_u \sum_{m=1}^{M} Y_m MW_m^{-1}$. $MW_m$ is the molar weight of $m$-th species and $R_u = 8.314$ J/(mol·K) is universal gas constant. The source terms in Eqs. (1)–(4), i.e. $S_{mass}$, $\mathbf{S}_{mom}$, $S_{energy}$ and $S_{species,m}$, account for the exchanges of mass, momentum, energy and species. Their corresponding expressions are given in Eqs. (26)–(29), respectively.

The viscous stress tensor $\mathbf{T}$ in Eq. (2) modelled by

$$\mathbf{T} = -2\mu \text{dev}(\mathbf{D}). \quad (6)$$

Here $\mu$ is dynamic viscosity and is dependent on gas temperature following the Sutherland's law [27]. Moreover, $\mathbf{D} \equiv [\nabla \mathbf{u} + (\nabla \mathbf{u})^T]/2$ is the deformation gradient tensor and its deviatoric component in Eq. (6), i.e. dev($\mathbf{D}$), is defined as dev($\mathbf{D}$) $\equiv \mathbf{D} - \text{tr}(\mathbf{D})\mathbf{I}/3$ with $\mathbf{I}$ being the unit tensor.

In addition, $\mathbf{j}$ in Eq. (3) is the diffusive heat flux and can be represented by Fourier's law, i.e.

$$\mathbf{j} = -k \nabla T. \quad (7)$$

Thermal conductivity $k$ is calculated using the Eucken approximation [28], i.e. $k = \mu C_v(1.32 + 1.37 \cdot R/C_v)$, where $C_v$ is the heat capacity at constant volume and derived from $C_v = C_p - R$. Here $C_p = \sum_{m=1}^{M} Y_m C_{p,m}$ is the heat capacity at constant pressure, and $C_{p,m}$ is the heat capacity at constant pressure of $m$-th species, which is estimated from JANAF polynomials [29].

In Eq. (4), $\mathbf{s_m} = -D_m \nabla(\rho Y_m)$ is the species mass flux. The mass diffusivity $D_m$ can be derived from heat diffusivity $\alpha = k/\rho C_p$ through $D_m = \alpha/Le_m$. With the unity Lewis number assumption (i.e. $Le_m = 1$), the mass diffusivity $D_m$ is calculated through $D_m = k/\rho C_p$. Moreover, $\dot{\omega}_m$ is the production or consumption rate of $m$-th species by all $N$ reactions, and can be calculated from the reaction rate of each reaction $\omega_{m,j}^o$, i.e.

$$\dot{\omega}_m = MW_m \sum_{j=1}^{N} \omega_{m,j}^o. \tag{8}$$

Also, the term $\dot{\omega}_T$ in Eq. (3) accounts for the heat release from chemical reactions and is estimated as $\dot{\omega}_T = -\sum_{m=1}^{M} \dot{\omega}_m \Delta h_{f,m}^o$. Here $\Delta h_{f,m}^o$ is the formation enthalpy of $m$-th species.

## *2.2 Governing equation for liquid phase*

The Lagrangian method is used to model the dispersed liquid phase, which is composed of a large number of spherical droplets [30]. The interactions between the droplets are neglected because we only study the dilute sprays with the droplet volume fraction being generally less than 1‰ [31]. The droplet break-up is not considered in this work. Therefore, the governing equations of mass, momentum and energy for the individual droplets are

$$\frac{dm_d}{dt} = -\dot{m}_d, \tag{9}$$

$$\frac{d\mathbf{u}_d}{dt} = \frac{\mathbf{F}_d}{m_d}, \tag{10}$$

$$c_{p,d} \frac{dT_d}{dt} = \frac{\dot{Q}_c + \dot{Q}_{lat}}{m_d}, \tag{11}$$

where $m_d$ is the mass of a single droplet and it is calculated by $m_d = \pi \rho_d d_d^3/6$ ($\rho_d$ and $d_d$ are the material density and diameter of a single droplet, respectively). $\mathbf{u}_d$ is the droplet velocity vector, $\mathbf{F}_d$ is the force exerting on the droplet and here we only consider the Stokes drag [32]. $c_{p,d}$ is the droplet heat capacity at constant pressure, and $T_d$ is the droplet temperature. In this work, both $\rho_d$ and $c_{p,d}$ are functions of droplet temperature $T_d$ [33], i.e.

$$\rho_d(T_d) = \frac{a_1}{a_2^{1+(1-T_d/a_3)^{a_4}}}, \tag{12}$$

$$c_{p,d}(T_d) = \frac{b_1^2}{\tau} + b_2 - \tau\left\{2.0 b_1 b_3 + \tau\left\{b_1 b_4 + \tau\left[\frac{1}{3}b_3^2 + \tau\left(\frac{1}{2}b_3 b_4 + \frac{1}{5}\tau b_4^2\right)\right]\right\}\right\}, \tag{13}$$

where $a_i$ and $b_i$ denote the species-specific constants and can be found from Ref. [33]. In Eq. (13), $\tau = 1.0 - min(T_d, T_c)/T_c$, where $T_c$ is the critical temperature (i.e. the temperature at and above which vapor of the substance cannot be liquefied, irrespective of the pressure) and $min\,(\cdot,\cdot)$ is the minimum function.

The evaporation rate, $\dot{m}_d$, in Eq. (9) is modelled through

$$\dot{m}_d = \pi d_d Sh D_{ab} \rho_s ln(1 + X_r), \tag{14}$$

where $X_r$ is the molar ratio, and estimated from

$$X_r = \frac{X_S - X_C}{1 - X_S}. \tag{15}$$

Here $X_C$ is the condensed species molar fraction in the surrounding gas, and $X_S$ is the fuel species molar fraction at the droplet surface. Note that estimation of the molar ratio in Eq. (15) does not account for the actual chemical reaction effects in the bulk gas. $X_S$ can be calculated using Raoult's Law

$$X_S = X_m \frac{p_{sat}}{p}, \tag{16}$$

with which it has been assumed the inter-molecular force difference in the mixture is neglected. This is expected to be acceptable for high-speed flows. $p_{sat}$ is the saturated pressure and is a function of droplet temperature $T_d$ [33], i.e.

$$p_{sat} = p \cdot exp\left(c_1 + \frac{c_2}{T_d} + c_3 ln T_d + c_4 T_d^{c_5}\right), \tag{17}$$

where $c_i$ are constants and can be found from Ref. [33]. Variations of $p_{sat}$ with $T_d$ is expected to accurately reflect the liquid droplet evaporation in high-speed hot atmosphere. Moreover, in Eq. (16), $X_m$ is the molar fraction of the condensed species in the gas phase. Moreover, in Eq. (14), $\rho_S = p_S MW_m/RT_S$ is vapor density at the droplet surface, where $p_S$ is surface vapor pressure and $T_S = (T + 2T_d)/3$ is droplet surface temperature. $D_{ab}$ is the vapor diffusivity in the gaseous mixture [27], i.e.

$$D_{ab} = 3.6059e^{-3} \times (1.8T_S)^{1.75} \times \frac{\alpha_l}{p_S \beta_l}, \tag{18}$$

where $\alpha_l$ and $\beta_l$ are constants related to different species.

The Sherwood number in Eq. (14) is modelled as [34]

$$Sh = 2.0 + 0.6 Re_d^{1/2} Sc^{1/3}, \tag{19}$$

where $Sc$ is the Schmidt number of gas phase and its value is assumed to be 1.0. The droplet Reynolds number in Eq. (19), $Re_d$, is defined based on the velocity difference between the two phases, i.e.

$$Re_d \equiv \frac{\rho_d d_d |\mathbf{u}_d - \mathbf{u}|}{\mu}. \tag{20}$$

The Stokes drag in Eq. (10) is modeled as (assuming that the droplet is spherical) [32]

$$\mathbf{F}_d = \frac{18\mu}{\rho_d d_d^2} \frac{C_d Re_d}{24} m_d (\mathbf{u} - \mathbf{u}_d). \tag{21}$$

The drag coefficient in Eq. (21), $C_d$, is estimated as [32]

$$C_d = \begin{cases} 0.424, & Re_d > 1000, \\ \frac{24}{Re_d}\left(1 + \frac{1}{6} Re_d^{2/3}\right), & Re_d < 1000. \end{cases} \tag{22}$$

The convective heat transfer rate $\dot{Q}_c$ in Eq. (11) is calculated by

$$\dot{Q}_c = h_c A_d (T - T_d), \tag{23}$$

where $A_d$ is the surface area of a single droplet. $h_c$ is the convective heat transfer coefficient, and computed using the correlation by Ranz and Marshall [34], i.e.

$$Nu = h_c \frac{d_d}{k} = 2.0 + 0.6 Re_d^{1/2} Pr^{1/3}, \tag{24}$$

where $Nu$ and $Pr$ are Nusselt and Prandtl numbers of gas phase, respectively.

The latent heat of vaporization $\dot{Q}_{lat}$ in Eq. (11) is calculated by

$$\dot{Q}_{lat} = h_{g,boil} - h_{l,boil}, \tag{25}$$

where $h_{g,boil}$ and $h_{l,boil}$ are the enthalpies for carrier gas phase and liquid phase respectively under boiling temperature.

Two-way coupling between the gas and liquid phases is taken into consideration. The corresponding terms, $S_{mass}$, $\mathbf{S}_{mom}$, $S_{energy}$ and $S_{species,m}$ in Eqs. (1)−(4), are calculated based on the contributions from each droplet in the host CFD cells, which read ($V_c$ is the cell volume, $N_d$ is the droplet number in the cell)

$$S_{mass} = \frac{1}{V_c} \sum_1^{N_d} \dot{m}_d, \tag{26}$$

$$\mathbf{S}_{mom} = -\frac{1}{V_c} \sum_1^{N_d} \mathbf{F}_d, \tag{27}$$

$$S_{energy} = -\frac{1}{V_c} \sum_1^{N_d} (\dot{Q}_c + \dot{Q}_{lat}), \tag{28}$$

$$S_{species,m} = \begin{cases} S_{mass} & for\ condensed\ species \\ 0 & for\ other\ species. \end{cases} \tag{29}$$

*2.3 Computational method*

The gas phase governing equations, i.e. Eqs. (1)−(4), are discretized with cell-centred finite volume method and solved by a density-based solver, *RYrhoCentralFoam*, which is developed from the fully compressible flow solver *rhoCentralFoam* in the framework of OpenFOAM 5.0 [35]. This solver can simulate compressible reactive flows and capture shock wave accurately in a collocated, polyhedral, finite-volume framework using non-oscillatory upwind-central schemes. The solver *rhoCentralFoam* is validated by Greenshields et al. [36] using various benchmark tests, including the Sod's problem, two-dimensional forward-facing step and supersonic jet. It is found that the central-upwind scheme in *rhoCentralFoam* (i.e. Kurganov, Noelle and Petrova (KNP) scheme [37]) is competitive with the state-of-the-art schemes, such as Roe scheme, in shock capturing [36]. The *RYrhoCentralFoam* solver has been systematically validated by a series of benchmark tests, including two-phase gas-droplet mixture [40]. The results show that the *RYrhoCentralFoam* solver can accurately predict the shock wave, molecular diffusion, auto-ignition and shock-induced ignition. Moreover, the sub-models of this solver related to the dispersed droplet phase are verified and validated against analytical and experimental data. It is also found that the *RYrhoCentralFoam* solver is able to model the gas−droplet two-phase detonations, including detonation propagation speed, interphase interactions and detonation frontal structures. The *RYrhoCentralFoam* solver has been applied for various detonative or RDC problems, including instability and wave bifurcation in RDC [10, 38], as well as two-phase RDC with liquid *n*-heptane [26]. Besides, the similar solver developed from *rhoCentralFoam* has also been validated and used in one- and two-dimensional detonations by Gutierrez Marcantoni et al. [39].

A second-order implicit backward method is employed for temporal discretization and the time step is about $10^{-9}$ s (maximum Courant number < 0.1). Second-order Godunov-type upwind-central KNP scheme [37] is used for discretizing the convective terms in momentum equation, i.e. Eq. (2). To ensure the numerical stability, van Leer limiter is adopted for flux calculations with KNP scheme. Total Variation Diminishing (TVD) scheme is used for the convective terms in the energy and species mass fraction equations. Also, second-order central differencing scheme is applied for the diffusion terms.

The chemical source terms, $\dot{\omega}_m$ and $\dot{\omega}_T$ in Eqs. (3) and (4) respectively, are integrated with a Euler implicit method. One-step irreversible mechanisms for $H_2$ [41] and $n$-$C_7H_{16}$ [42] combustion are used to estimate the reaction rates through

$$\omega_{m,j}^o = AT^n exp\left(-\frac{E_a}{RT}\right)[F]^a[O]^b, \tag{30}$$

where *A* is the pre-exponential factor and *n* is the temperature exponent. [*F*] and [*O*] are the concentrations of the fuel and oxidizer, respectively, $E_a$ is the activation energy, *a* and *b* are the fuel and oxidizer reaction orders, respectively. The kinetic parameters of the chemical reactions for $H_2/O_2$ and $n$-$C_7H_{16}/O_2$ are listed in Table 1.

Table 1. Chemical reactions for $H_2/O_2$ and $n$-$C_7H_{16}/O_2$ (units in cm-sec-mole-cal-Kelvins)

| Index | Reaction | $A$ | $n$ | $E_a$ | $a$ | $b$ | Ref. |
|---|---|---|---|---|---|---|---|
| I | $2H_2 + O_2 \Rightarrow 2H_2O$ | $1.03 \times 10^9$ | 0.0 | 29,841.0 | 1.0 | 1.0 | [41] |
| II | $n\text{-}C_7H_{16} + 11O_2 \Rightarrow 7CO_2 + 8H_2O$ | $5.10 \times 10^{11}$ | 0.0 | 30,000.0 | 0.25 | 1.5 | [42] |

The kinetic parameters of Reaction I in Table 1 are validated by Ma et al. [41] with one-dimensional detonation initiation / propagation, and successfully used in their simulations of pulse detonation engines. It is also used by Meng et al. [43] for modelling rotating detonation combustion and reasonable results of RDC propagation are achieved. Reaction II has been used for modelling spray RDC by Zhao and Zhang [26] and the results with Reaction II demonstrate good agreements with those from a skeletal $n$-$C_7H_{16}$ mechanism [44]. Since this study is focused on the RDW propagation and droplet dynamics, instead of detailed gaseous reactions, the foregoing mechanisms are sufficient. One-step or simplified mechanisms are also used for two-phase RDC modelling [9, 25] and general two-phase detonation problems [45, 46].

For the liquid phase, their equations, i.e. Eqs. (9)−(11), are solved by first-order implicit Euler method. Meanwhile, the gas properties at the droplet location are estimated based on linear interpolation. The two-way coupling terms, i.e. Eqs. (26) −(29), are calculated explicitly, as the source terms of the gas phase equations, i.e. Eqs. (1)−(4).

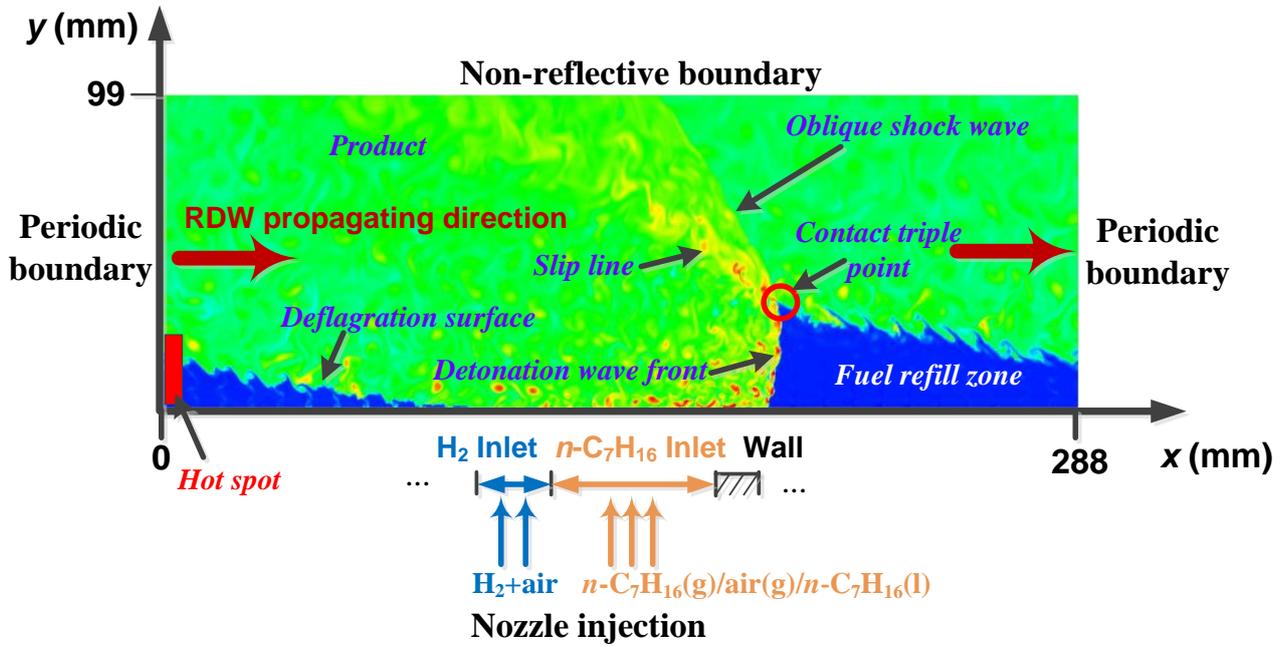

Fig. 1 Two-dimensional RDC model: computational domain and boundary condition. The gas temperature (250−3,000 K) is shown as the background.

## 3. Physical model and simulation case

### 3.1 RDC model with liquid fuel sprays

Figure 1 shows a two-dimensional rectangular domain which is used here as a simplified model of a flattened RDE combustor. The length (i.e. $x$-direction) of the computational domain in Fig. 1 is 288 mm, while the height ($y$-direction) is 99 mm. This domain has been used in our previous work by Zhao and Zhang [26], in which successful rotating detonation wave propagation in liquid $n$-$C_7H_{16}$ sprays is achieved. The corresponding nominal diameter is approximately 92 mm, similar to that of a small-scale RDE combustors (e.g. by Bohon et al. [47] and also Deng et al. [48, 49]). It should be highlighted that, to obtain physically sound results on RDC's with droplet evaporation and reactant mixing, it is significant to use a practically relevant dimensions for the simplified RDC model, in which the timescale relations between droplet evaporation, reactant mixing and periodic RDW propagation can be reasonably reproduced. In addition, turbulence effects on RDW is not considered here.

The boundary conditions of the RDC model are marked in Fig. 1. The outlet is assumed to be non-reflective, to avoid the interference from upstream-propagating pressure waves on the detonation wave and deflagration surface near the head end. The detonation waves can propagate continuously via periodic boundaries at the left and right sides of the domain, as indicated in Fig. 1. The modelled propellant injectors lie at the entire lower boundary in Fig. 1, which will be introduced in Section 3.2.

The RDW is initiated forcibly using a rectangular hot spot (20 atm and 2,000 K) near the left boundary with height and width being 12.2 mm × 1 mm (see Fig. 1), respectively. Additionally, a fresh fuel layer (12.2 mm × 288 mm) with premixed stoichiometric $n$-$C_7H_{16}$/air is initialized near the lower boundary (i.e. the fuel injectors) in Fig. 1 to promote the propagation of the newly ignited RDW. The rest region in the domain is initialized with air, corresponding to pressure of 1 atm and temperature of 300 K. Stable propagation of RDW is identified through examining the pressure history from probes and instantaneous wave propagation speed.

*3.2 Gaseous / liquid reactants and injection model*

In this work, the fuels are partially pre-vaporized $n$-$C_7H_{16}$ sprays with separate injection of stoichiometric hydrogen/air gas. The latter is used for stabilizing the continuous detonations, as implemented in the liquid fuel RDE experiments by Bykovskii et al. [19, 20] and Kindracki [22]. The reactant injector system is modelled as 32 sets of abreast and continuously arranged $n$-$C_7H_{16}$ and $H_2$ inlet jets at the head end (see Fig. 1). A set of these two inlets are discretized with solid walls to mimic the discrete injection of propellants in practical RDE's. The area ratio $\theta$ (length ratio in 2D cases) between $H_2$, $n$-$C_7H_{16}$ inlets and the solid wall is fixed to be 1:6:2. Although it is shown that the injector spacing may affect the RDW propagation [9, 25], however, this effect is not studied in this work.

Heterogeneous mixtures of $n$-$C_7H_{16}$ droplets and vapour are injected through the $n$-$C_7H_{16}$ inlet. The droplets are assumed to be spherical and mono-sized, with diameters being 5−50 μm, which are close to those measured in detonation combustor by Kindracki [21]. It is acknowledged that polydispersity of fuel droplets is not considered. Computational parcel method is used in our simulations, in which droplets with identical properties are denoted by one parcel. In addition, different levels of pre-vaporization in the $n$-$C_7H_{16}$ stream are parameterized by the equivalence ratio $\phi_g$ of premixed vapour/air mixture through the $n$-$C_7H_{16}$ inlet. Both gaseous $n$-$C_7H_{16}$ and $H_2$ streams have identical total temperature and pressure, i.e. $T_0$ = 300 K and $p_0$ = 20 atm respectively.

Similar to the gaseous fuel RDE modelling [11, 50-54], injections of $n$-$C_7H_{16}$ and/or $H_2$ are determined by the relation between injector total pressure and local pressure near the inlet in the domain. For the liquid phase, the interphase kinematic equilibrium is assumed, and hence the droplet injection velocity follows that of the gas phase issued from the same inlets. This assumption is acceptable when the droplets disperse sufficiently long in the carrier gas from the upstream plenum of practical RDE's. Therefore, in our studies the $n$-$C_7H_{16}$ spray injection is activated if, and only if, the top head gas pressure is lower than $p_o$. With the above assumptions and simplifications, there are three situations for $n$-$C_7H_{16}$ inlets, i.e.

(1) If $p \geq p_o$, then there are no reactants injected through the inlets, which are treated as solid walls and hence no flashback can occur. The local pressure $p_i$, gas temperature $T_i$, droplet temperature $T_{i,d}$, gas velocity $v$ and droplet velocity $v_d$ respectively satisfy the following conditions ($T$ is the local temperature near the inlet)

$$p_i = p, T_i = T \text{ and } v = 0, \text{ no droplets}; \tag{31}$$

(2) If $p_{cr} < p < p_o$, then the flows at the inlet are not choked, and therefore

$$p_i = p, \ T_i = T_o \left(\frac{p_i}{P_o}\right)^{\frac{\gamma_R-1}{\gamma_R}}, \ T_{i,d} = T_o, \text{ and } v = v_d = \sqrt{\frac{2\gamma_R}{\gamma_R-1}RT_o\left[1 - \left(\frac{p_i}{P_o}\right)^{\frac{\gamma_R-1}{\gamma_R}}\right]}; \tag{32}$$

(3) If $p \leq p_{cr}$, then the flows at the inlet are choked, and therefore

$$p_i = p_{cr}, \ T_i = T_o \left(\frac{p_i}{P_o}\right)^{\frac{\gamma_R-1}{\gamma_R}}, \ T_{i,d} = T_o, \text{ and } v = v_d = \sqrt{\frac{2\gamma_R}{\gamma_R-1}RT_o\left[1 - \left(\frac{p_i}{P_o}\right)^{\frac{\gamma_R-1}{\gamma_R}}\right]}. \tag{33}$$

The critical pressure $p_{cr}$ is calculated from the choking condition, based on the total pressure and reactant specific heat ratio, $p_{cr} = p_o \left(\frac{2}{\gamma_R+1}\right)^{\frac{\gamma_R}{\gamma_R-1}}$. $\gamma_R$ is the specific heat capacity ratio of the mixture. For H$_2$ inlets, same correlations as Eqs. (31)−(33) are used, without specifications of $T_{i,d}$ and $v_d$. A *posterior* examination of our results show that this droplet injection model can accurately reproduce the specified droplet mass flow rates (with errors less than 2%), and it has been successfully applied in our recent simulations of two-phase RDE's [26].

Table 2. Information of liquid and pre-vaporized *n*-heptane (stoichiometric H$_2$/air from H$_2$ inlet)

| Case group | *n*-Heptane equivalence ratio | | | Hydrogen | Liquid Droplet property | |
|---|---|---|---|---|---|---|
| | Liquid *n*-heptane phase $\phi_l$ | Pre-vaporized *n*-heptane gas phase $\phi_g$ | Global *n*-heptane Equivalence ratio† $\phi_t$ | Mass flow rate (kg/s) | Mass flow rate (kg/s) | Initial diameter $d_d^0$ (μm) |
| A | 0.134 | 0.35 | 0.446 | 0.037 | 0.107 | 5−50 |
| B | 0.129 | 0.4 | 0.485 | 0.036 | 0.097 | |
| C | 0.126 | 0.6 | 0.650 | 0.036 | 0.092 | |
| D | 0.123 | 0.8 | 0.813 | 0.035 | 0.087 | |
| E | 0.123 | 1.0 | 0.976 | 0.035 | 0.087 | |

† $\phi_t$ is estimated based on liquid / gaseous *n*-C$_7$H$_{16}$ and oxygen from both *n*-C$_7$H$_{16}$ and H$_2$ inlets.

*3.3 Simulation case*

Table 2 shows five case groups, i.e. A−E, simulated in this work. They are characterized by the different equivalence ratios of gas and liquid phases in the *n*-C$_7$H$_{16}$ stream, i.e. $\phi_g$ and $\phi_l$. They are respectively estimated based on the mass of the gaseous and liquid *n*-heptane and the air injected from the *n*-heptane inlet. As shown in Table 2, the global *n*-C$_7$H$_{16}$ equivalence ratio $\phi_t$ gradually increases, from Case A with $\phi_t = 0.446$ to Case E with $\phi_t = 0.976$. In each group, the droplet diameters considered are $d_d^0$ = 5, 10, 15, 20, 30, 40 and 50 μm. Readers should be reminded that in all the cases, the hydrogen/air stream is stoichiometric. For ease of reference, individual cases are named as, e.g. A15. Here the letter A denotes case group A with $\phi_l = 0.134$ and $\phi_g = 0.35$, whereas the number indicates the initial droplet diameter, i.e. $d_d^0 = 15$ μm. For each group, the corresponding droplet-free RDC cases are also simulated for comparisons. For instance, A00 denotes the purely gaseous *n*-heptane/air mixture with $\phi_g = 0.35$.

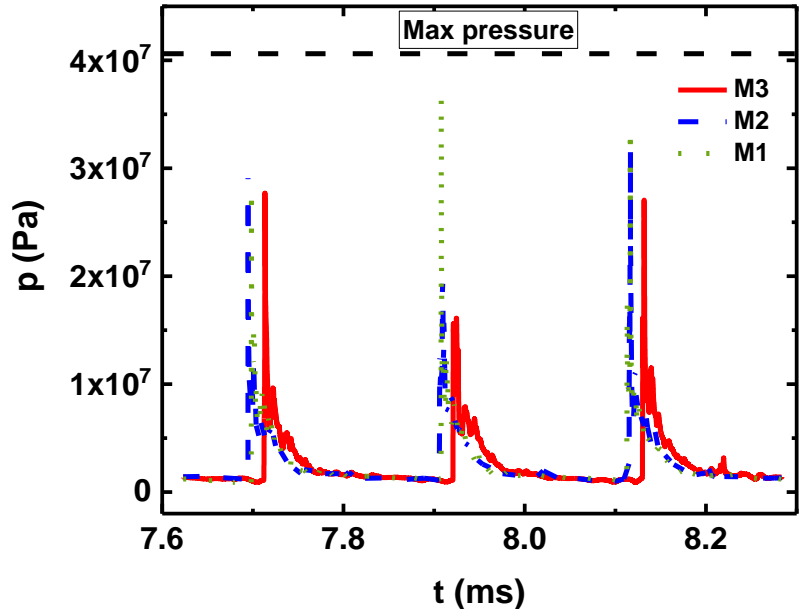

Fig. 2 Pressure history from a location near the head end, Case B15. Dashed line: maximum pressure from SD Toolbox [55].

## 4. Mesh sensitivity analysis

The computational domain in Fig. 1 is discretized with uniform Cartesian cells. It should be highlighted that resolving the detailed detonation structure is beyond the objective of this work; instead, detonation propagation in two-phase fuels are concentrated. Meanwhile, in Lagrangian tracking of the sub-grid droplets, a mesh resolution larger than the droplet diameter is desirable, to ensure that the gas phase quantities at the droplet surface (critical for estimating the two-phase coupling, e.g. evaporation) can be accurately approximated using the interpolated ones at the droplet centre [56]. Therefore, three mesh sizes are tested, i.e. 0.1, 0.2 and 0.4 mm, termed as M1, M2 and M3, respectively.

Figure 2 shows the pressure history at a probe near the inlet end with M1, M2 and M3. The results are from Case B15. It is found that variations of the mesh resolutions do not change the RDW number and propagation direction. Also, from Fig. 2, the pressure histories predicted with M1 and M2 are close, in terms of the peak pressure instants and values. However, pressure history from M3 slightly lags behind those from M1 and M2 and the peak value is also slightly lower than those from M1 and M2. The higher pressure from M1 at $t = 7.9$ s may result from the non-uniform instantaneous pressure distribution along the detonation wave. Moreover, the pressure peaks from M1 and M2 are closer to the maximum pressure ($4.09 \times 10^7$ Pa, predicted by SD Toolbox [55] with 250 K and 15 atm).

Table 3 RDW propagation speeds predicted with different meshes

| Mesh | $\Delta_{x,y}$ (mm) | $D_{det}$ (m/s) | $D_{CJ}$ (m/s) | Deficit (C–J, %) | Deviation (M1, %) |
|---|---|---|---|---|---|
| M1 | 0.1 | 1,379 | 1,615 | 14.6 | – |
| M2 | 0.2 | 1,386 | | 14.2 | 0.5 |
| M3 | 0.4 | 1,359 | | 15.8 | 1.5 |

Table 3 shows the RDW propagation speed $D_{det}$ predicted with the three meshes. $D_{CJ}$ is also calculated with SD Toolbox based on the assumption that all the $n$-$C_7H_{16}$ droplets are vaporized and perfectly mixed with the $n$-$C_7H_{16}$ and $H_2$ gaseous mixtures. The deficits of the calculated detonation speed relative to Chapman–Jouguet (C–J) speed are all around 15%, which may stem from the dual-fuel and two-phase injections, and hence non-uniform reactant distributions. Moreover, the deviations of RDW propagation speed relative to the finest case M1 increase with increased mesh size, 1.5% for M3 and 0.5% for M2.

Two-way interphase coupling is considered in this work and therefore it is also important to examine how mesh resolutions for Eulerian gas phase influence the Lagrangian droplet calculations. Figure 3 shows the variations of droplet arithmetic mean diameter ($d_{10}$) along the height ($y$) direction in Case B15. Here $d_{10}$ is estimated based on the droplets in the entire domain from about 50 instants.

The droplet diameters from M3 are consistently higher than the results in M1 and M2. Meanwhile, the droplet distributions in M1 and M2 are generally similar along the entire RDE height direction. Based on the results in Figs. 2 and 3 as well as Table 3, the resolution of 0.2 mm will be selected, in terms of the acceptable computational accuracy and cost. Similar resolutions are also used by Schwer et al. [45, 57] for their Eulerian−Lagrangian modelling of two-phase detonation.

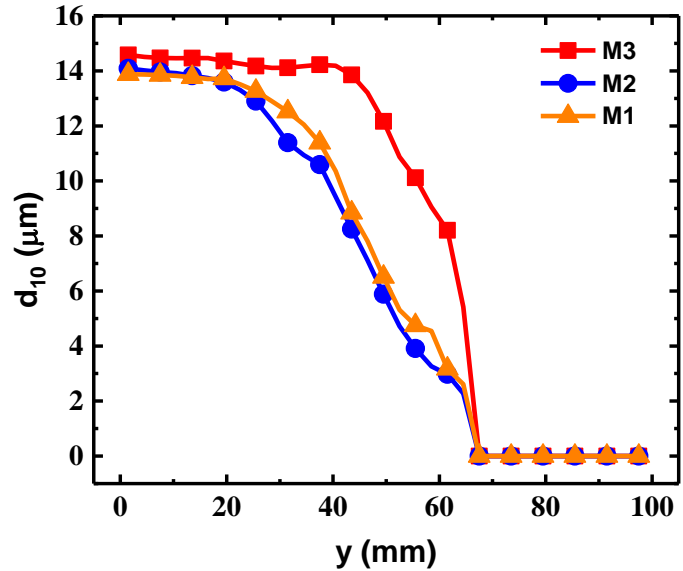

Fig. 3 Distributions of mean droplet diameter along the RDE height, Case B15.

## 5. Results and discussion

### *5.1 General characteristics of liquid-fuelled rotating detonation combustion*

Figure 4 respectively demonstrate the distributions of Lagrangian droplets, gas temperature (colorcut off over 310 K), pressure, averaged evaporation rate, mass fractions of *n*-heptane and hydrogen from a single-waved case, i.e. B05. Note that the averaged evaporation rate is calculated using Eq. (26). Figures 4(d) and 4(e) respectively visualize the mass fractions of *n*-$C_7H_{16}$ vaporized in the chamber and pre-vaporized before injection. The details of the droplets inside the dashed box in Fig. 4(a) are enlarged in Fig. 5.

It can be found from Figs. 4(a) and 4(b) that the RDC structure includes detonation wave, deflagration surface and oblique shock wave, similar to the previously results of two-phase [9, 25, 26] and gaseous RDC [10, 52]. The detonation wave height (*y*-direction) $h_d$ is 0.03 m, whilst the refill zone length (x-direction) $l_f$ is about 0.193 m. Their ratio, $K = l_f/h_d$, is 6.4, which is within the range (i.e. 5−9) suggested by Bykovskii et al. [58] for gaseous reactants. The discrete inlets result in the ribbon-like distributions of *n*-heptane and hydrogen fuels in the refill zone. As mentioned in Section 3.2, the injections of liquid *n*-$C_7H_{16}$ droplets at the individual inlets are synchronized with that of the carrier gas. Therefore, droplet injection only occurs in the triangular fuel refill zone and leads to a series of parallelly spray jets, as shown by the Lagrangian droplets in Figs. 4(a) and 5(a). Behind the RDW, due to the higher local pressure than the total pressure (i.e. 20 atm), the droplet injection is suppressed.

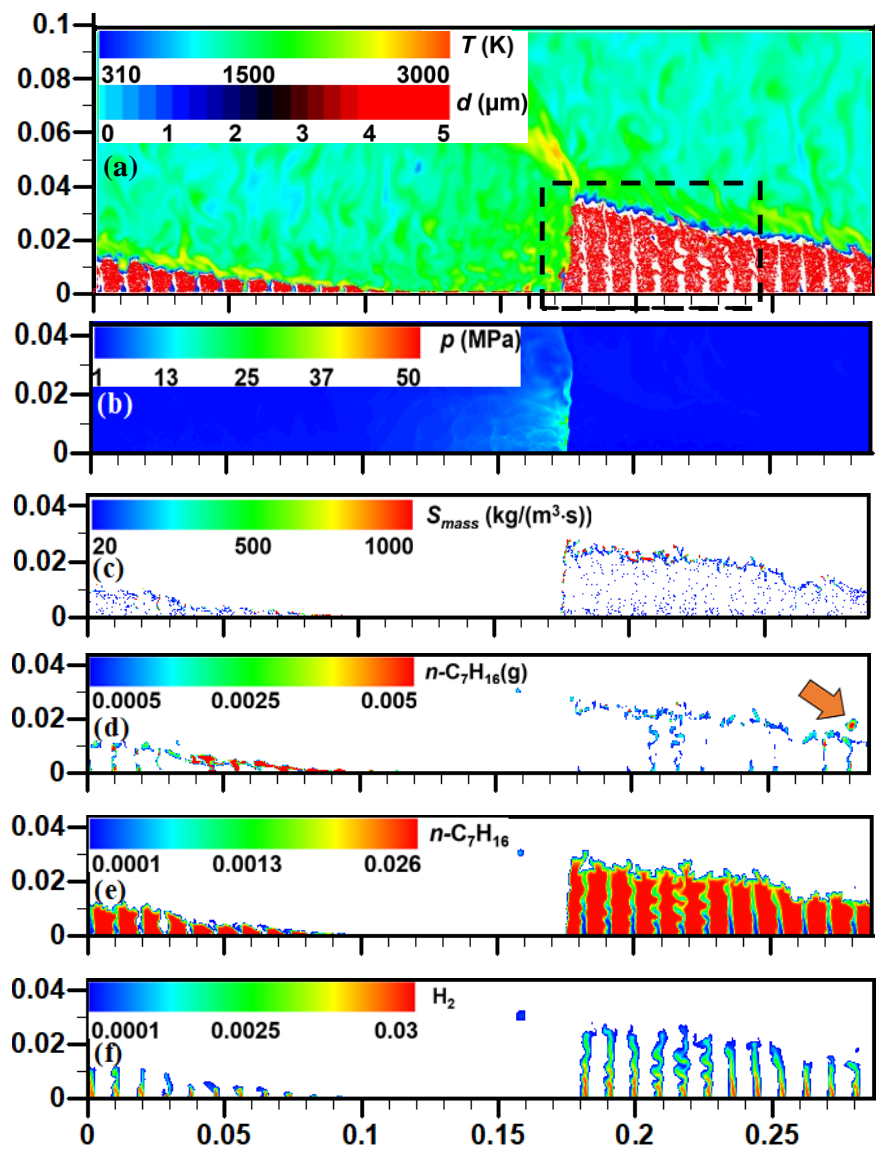

Fig. 4 Contours of (a) temperature (lower limit clipped to 310 K), (b) pressure, (c) averaged evaporation rate, (d) vaporized $n$-$C_7H_{16}$ from droplets, (e) pre-vaporized $n$-$C_7H_{16}$, (f) $H_2$. Results from Case B05. Unit for $x$- and $y$-axis in m.

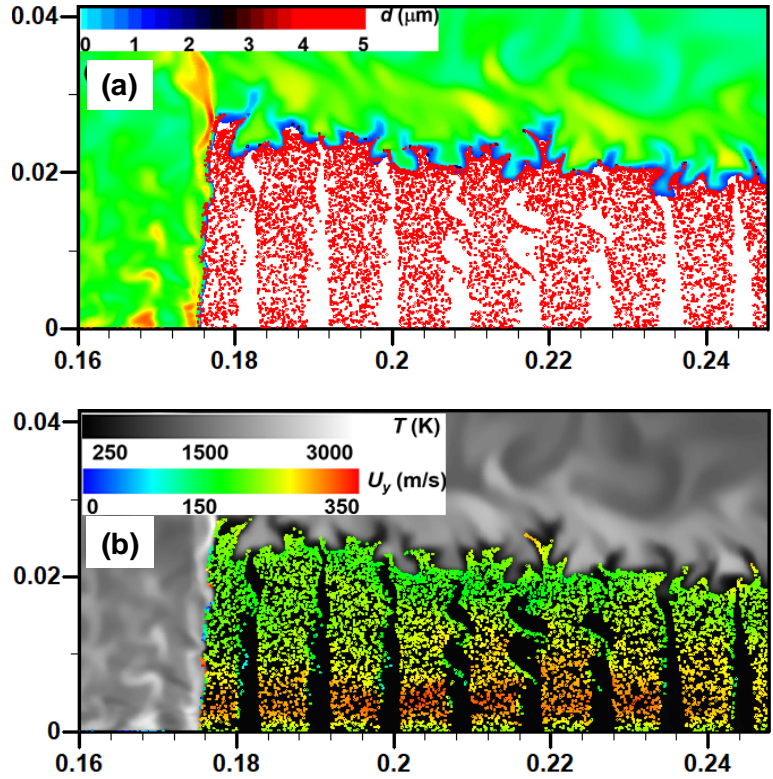

Fig. 5 Distributions of droplets coloured by (a) diameter and (b) $y$-direction velocity. The region corresponds to the dashed box in Fig. 4(a). Unit for $x$- and $y$-axis in m.

Droplet evaporation is limited inside the refill zone as shown from the distribution of averaged evaporation rate in Fig. 4(c). This can also be confirmed by negligible decrease of droplet diameters (see Figs. 4a and 5a). On the contrary, high evaporation rates are observable in Fig. 4(c) near the deflagration surface. Due to small diameters ($d_d^0 = 5$ μm) and large $y$-direction velocity (see Fig. 5b), the droplets are transported by the carrier gas, traversing the refill zone. For those droplets at the tips of the spray jets, they have direct contact with the wrinkled deflagration surface and evaporation occurs due to the local high temperature, which can be found in Fig. 5. Nevertheless, through comparisons between Figs. 4(d) and 4(e), one can see that the concentration of the pre-vaporized $n$- $C_7H_{16}$ is much higher than that of $n$-$C_7H_{16}$ vaporized in the chamber. Interestingly, the droplets do not penetrate into the burned side of the deflagration surface, since they have small velocity differences with the local surrounding gas and are engulfed and trapped by the eddies resulting from Kelvin–Helmholtz and Richtmyer–Meshkov instabilities [10, 53, 59] (see Fig. 5). However, a pocket of the vapour is occasionally transported into the burned gas side (pointed by the arrow in Fig. 4d) and then deflagrated there.

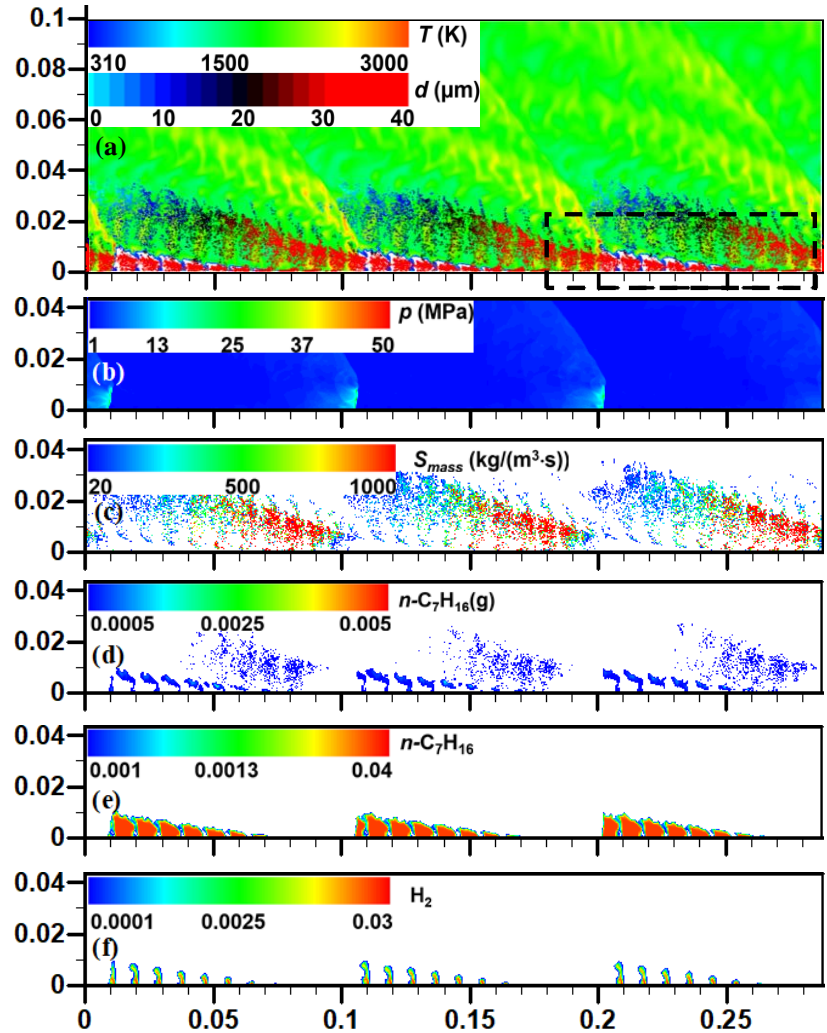

Fig. 6 Contours of (a) temperature (lower limit clipped to 310 K), (b) pressure, (c) evaporation rate, (d) vaporized $n$-$C_7H_{16}$ from droplets, (e) pre-vaporized $n$-$C_7H_{16}$, (f) $H_2$. Results from Case C40. Unit for $x$- and $y$-axis in m.

Plotted in Figs. 6 and 7 are the counterpart results from Case C40. Different from the results in Figs. 4 and 5, three RDW's are stabilized in this case. The bifurcation of the RDW number results from the spontaneous ignition of detonation wavelets along the deflagration surface of the refill zone. Three RDWs' have nearly the same height, i.e. $h_d \approx 0.01$ m, about one third of that in Fig. 4. Accordingly, the length of the triangular fuel refill zone is reduced to $l_f \approx 0.055$ m, leading to $K = 5.5$. In this case, a large number of liquid droplets can be seen behind the detonation waves. This may be caused by the shorter refill zone and increased initial droplet diameter ($d_d^0 = 40$ μm). These escaping droplets are accelerated by expanded post-detonation gas (characterized by the high velocity in the post-detonation areas in Fig. 7). Moreover, continuous evaporation occurs there due to the high gas temperature and/or gradually reduced pressure, as seen from the distributions of high evaporation rates in Fig. 6(c). This is consistent with the gradual reduction of the droplet diameters when the distance between the droplet and the detonation wave increases (see Fig. 7). It should be emphasized that droplets at two sides of the deflagration surfaces are originally different: they are newly injected in the refill zone and escape from the preceding detonation front, respectively. This is apparent by their distinct droplet diameters (see Fig. 7a).

It is noteworthy to further discuss the interactions between liquid droplets and the characteristic structures (e.g. detonation wave, deflagrative surface and oblique shock wave) in rotating detonation combustion, through the droplet distributions in Figs. 5 and 7. For small droplets, e.g. 5 μm in Fig. 5, the droplets around the approaching detonation wave are completely vaporized and the vapours are detonated. Meanwhile, the hot burned gas near the deflagrative surface promotes the evaporation there, leading to high vapour concentration and subsequent deflagrative combustion. The above phenomena are qualitatively consistent with the observations by Hayashi et al. [25] and Sun and Ma [9] in their RDC modelling with liquid JP-10 ($d_d^0 = 1, 5,$ and 10 μm) and octane sprays ($d_d^0 = 20$ μm), respectively. However, for relatively large droplets (e.g. 40 μm in Fig. 7b), they stagnate slightly away from the deflagration surface due to the local kinematic equilibrium effects. For those escaping droplets, when they cross the oblique shock wave extending from the next RDW's, most of them are depleted due to the high temperature in the post-shock area, as observed in Fig. 6(a). Representative droplet trajectories in an RDE chamber will be further studied in Section 5.4.

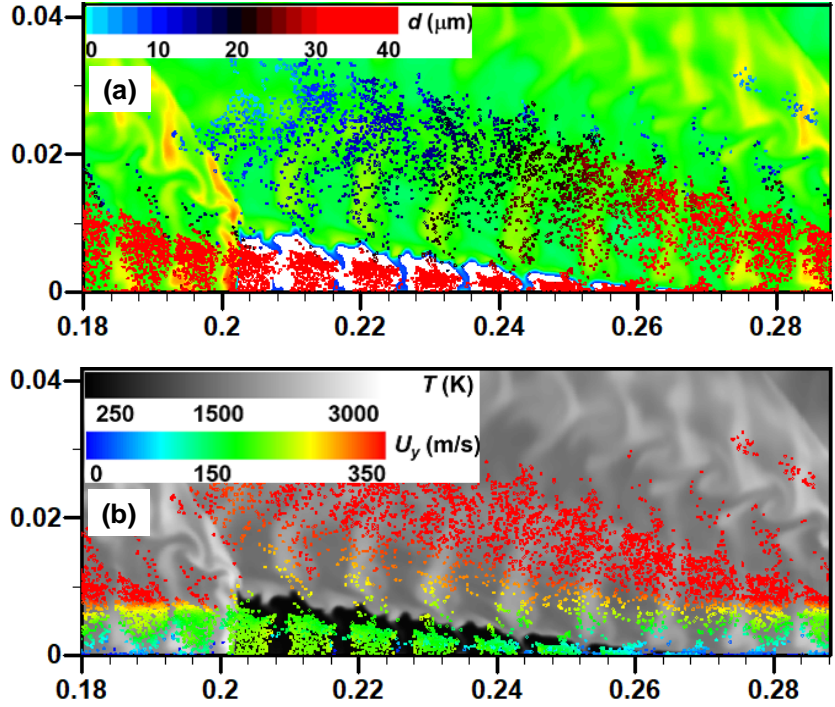

Fig. 7 Distributions of droplets coloured by (a) diameter and (b) *y*-direction velocity. The region corresponds to the dashed box in Fig. 6(a). Unit for *x*- and *y*-axis in m.

*5.2 Detonation propagation speed*

Figure 8 shows the detonation propagation speed $D_{det}$ and velocity deficit $D_f$ as functions of initial droplet size $d_d^0$ under various pre-vaporized gas equivalence ratios $\phi_g$. The solid lines are fitted from our simulation results marked as symbols. Note that each case group has a constant pre-vaporized gas equivalence ratio $\phi_g$ as tabulated in Table 2. It is seen that with increased $\phi_g$ from 0.35 to 1.0, $D_{det}$ increases. It is observed that for the relatively small $\phi_g$ (e.g. 0.35 and 0.4 in groups A and B), $D_{det}$ first increases as the droplet size increases, and when $d_d^0$ is beyond 5 μm, it gradually decreases and almost is constant (close to the corresponding gaseous $D_{det}$ values when their RDW numbers are the same) when $d_d^0 > 30$ μm. The speed enhancement for small droplets is probably due to their fast evaporation, kinetically contributing towards the overall stoichiometric gas composition. However, when $\phi_g$ further increases, e.g. 0.8 and 1.0 in groups C and D, such non-monotonicity becomes not obvious.

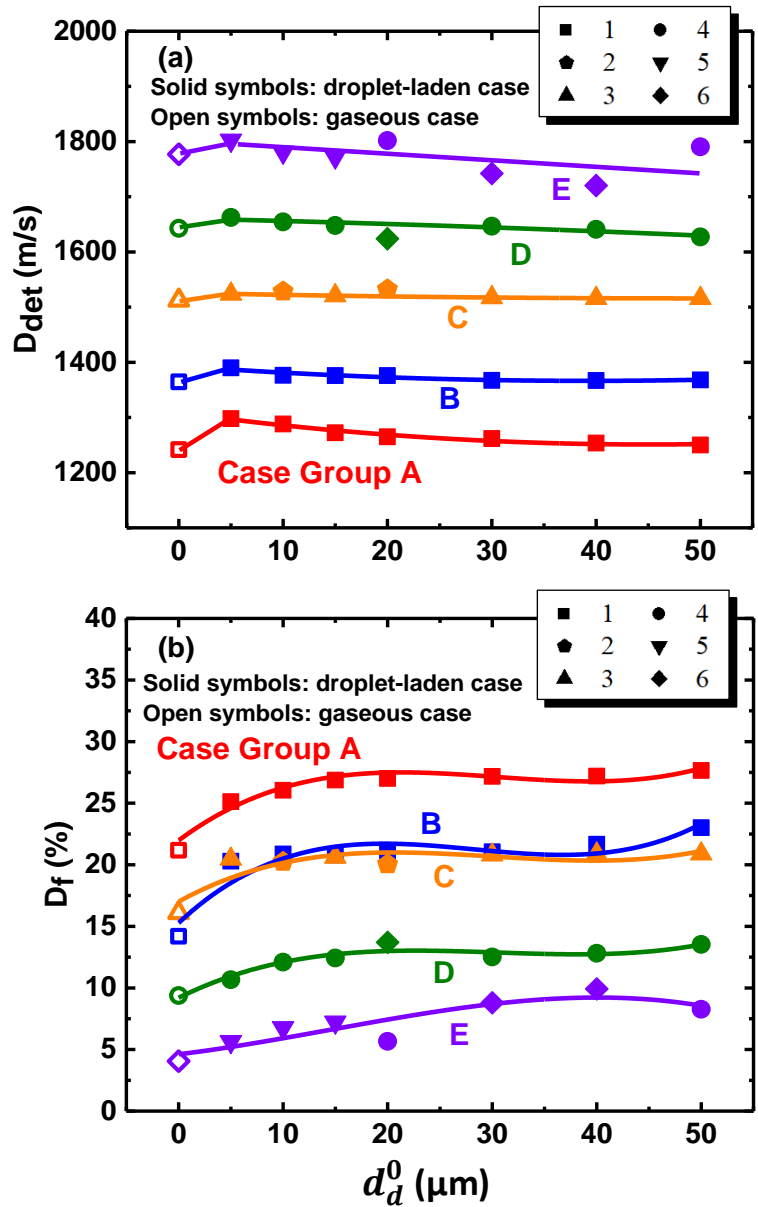

Fig. 8 (a) Detonation propagation speed and (b) velocity deficit as functions of initial droplet diameter under different pre-vaporized gas equivalence ratios. Open symbols: gaseous (i.e. $n$-$C_7H_{16}$ vapour and $H_2$) case without droplets injection. The number in the legend indicates the RDW number, which also applies for Figs. 9 and 21.

It is also seen from Fig. 8(a) that bifurcation of the RDW number occurs when $\phi_g$ is high, e.g. in groups C−E. Multiply RDW's are also observed by Bykovskii et al. [19] from the experimental studies of RDE with liquid kerosene−hydrogen−air mixtures, in which the RDW's increase from one to five with increased overall reactant mass flow rate. It is shown from Fig. 8(a) that the number of RDW's also affects the propagation speed of the two-phase detonation. This is particularly obvious in case group E. For example, detonation propagation speeds in Cases E20 and E50 (four-wave mode) are higher than those in adjacent cases (Cases E15 and E30, five- and six-wave modes respectively). Similar tendencies can also be observed among Cases D30, D20, D15. This is because as the RDW number increases, the width of the fuel refill zone decreases accordingly, the time for droplet evaporation and/or reactant mixing is reduced.

Figure 8(b) further shows the corresponding velocity deficits, which are relative to the respective C−J velocities calculated assuming that all the liquid $n$-$C_7H_{16}$ are fully vaporized and the gaseous mixture is perfectly mixed. Compared to the gaseous cases, the sprayed fuels lead to higher velocity deficits, and the differences are more striking when the pre-vaporization level is relatively small (e.g. $\phi_g$ = 0.35, 0.4 and 0.6). Meanwhile, for the same $\phi_g$, the velocity deficit $D_f$ generally rises when the initial droplet size increases, consistent with the tendencies from liquid-fuelled detonation in tubes [60, 61]. However, as $\phi_g$ increases, $D_f$ consistently decreases. For example, the velocity deficits of group E vary between 4%−10%, whilst those of group A 21%−28%. The exceptions are groups B and C. $D_f$ in these two groups are close because generally higher RDW number in group C than in group B, which to some degree offset the deficit reduction caused by increased pre-vaporization gas equivalence ratio. Similar velocity deficit level, i.e. 20%−25%, are reported by Kindracki [22] in his liquid kerosene RDE experiments. It is well known that there are diverse factors for velocity deficits in RDE's [1], and in our cases, they may be caused by droplet evaporation, gas reactant mixing and/or RDW number multiplicity.

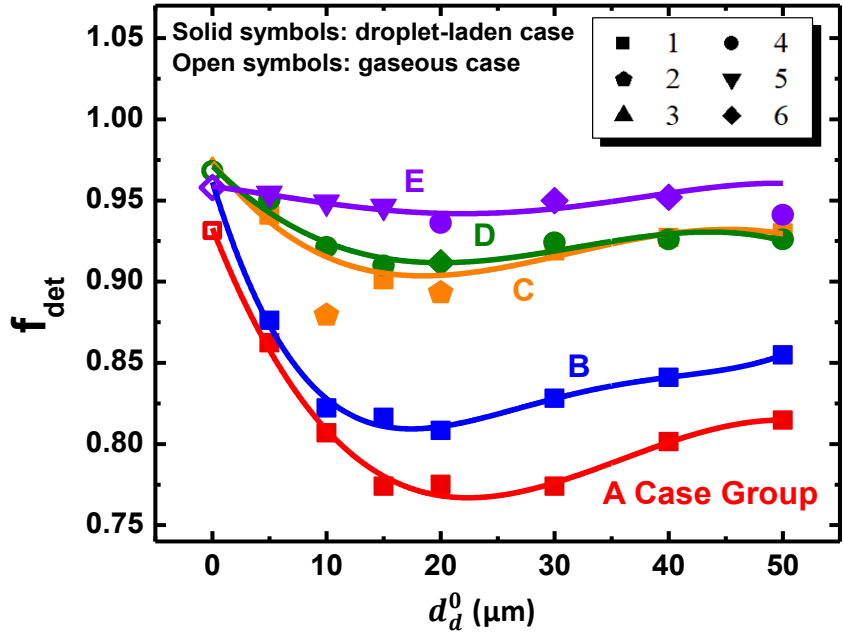

Fig. 9 Detonated fuel fraction as a function of initial droplet diameter under different pre-vaporized gas equivalence ratios.

*5.3 Detonated fuel fraction*

Figure 9 shows the fraction of detonated fuels (including $n$-$C_7H_{16}$ and $H_2$), $f_{det}$, as a function of initial droplet diameter $d_d^0$ under different gas equivalence ratios $\phi_g$. $f_{det}$ is estimated based on the volume averaged detonation consumption rates of individual fuel conditioning on heat release rate $\dot{Q}$ greater than $10^{13}$ J/m$^3$/s, approximately deem to be detonative combustion [10], i.e.

$$f_{det} = \frac{\dot{\omega}_{det,n-C7H16} + \dot{\omega}_{det,H2}}{\dot{\omega}_{n-C7H16} + \dot{\omega}_{H2}}, \tag{34}$$

where $\dot{\omega}_{n-C7H16}$ and $\dot{\omega}_{H2}$ are the $n$-$C_7H_{16}$ and $H_2$ overall consumption rates, respectively. $\dot{\omega}_{det,n-C7H16}$ and $\dot{\omega}_{det,H2}$ are the $n$-$C_7H_{16}$ and $H_2$ consumption rates, respectively, conditioning on $\dot{Q} > 10^{13}$ J/m$^3$/s. In general, for a fixed droplet diameter $d_d^0$, $f_{det}$ increases from groups A to E. This is because $n$-$C_7H_{16}$ concentration in the gas phase is increased with $\phi_g$. Meanwhile, this enhancement effect on detonative combustion are more pronounced when $\phi_g$ is relatively low, for instance, in groups A, B and C. There exists a critical droplet diameter, i.e. about 20 μm, corresponding to minimum $f_{det}$. When $d_d^0 < 20$ μm, $f_{det}$ decreases with $d_d^0$. This is because larger droplets have less propensity to be fully vaporized at the detonation wave (hence smaller $\dot{\omega}_{det,n-C7H16}$). Meanwhile, the escaping droplets can continue evaporation immediately behind the RDW, and the vapour is consumed by the deflagration (larger $\dot{\omega}_{n-C7H16}$). These two processes result in smaller $f_{det}$. However, when $d_d^0 > 20$ μm, droplets are vaporized in a broader zone, i.e. between the slip line and deflagration surface behind the RDW, as seen from Figs. 6(a) and 7. Limited deflagrative combustion is observed there, due to low oxygen concentration under initial fuel-lean $n$-C₇H₁₆/air conditions (e.g. groups A and B). As such, $f_{det}$ increases with further increased $d_d^0$, because $\dot{\omega}_{n-C7H16}$ is reduced. Due to the deflagrative combustion of part of the vaporized $n$-C₇H₁₆, $f_{det}$ in the two-phase cases is generally lower than that in the corresponding gaseous cases. When the equivalence ratio of the initial $n$-C₇H₁₆/air gas becomes higher (e.g. groups C−E), the dependence of $f_{det}$ on $d_d^0$ tends to be weak. This can be justified by the limited evaporation time in the reduced refill zone due to wave bifurcation, and low oxygen concentration in the detonated gas due to the near-stoichiometric condition.

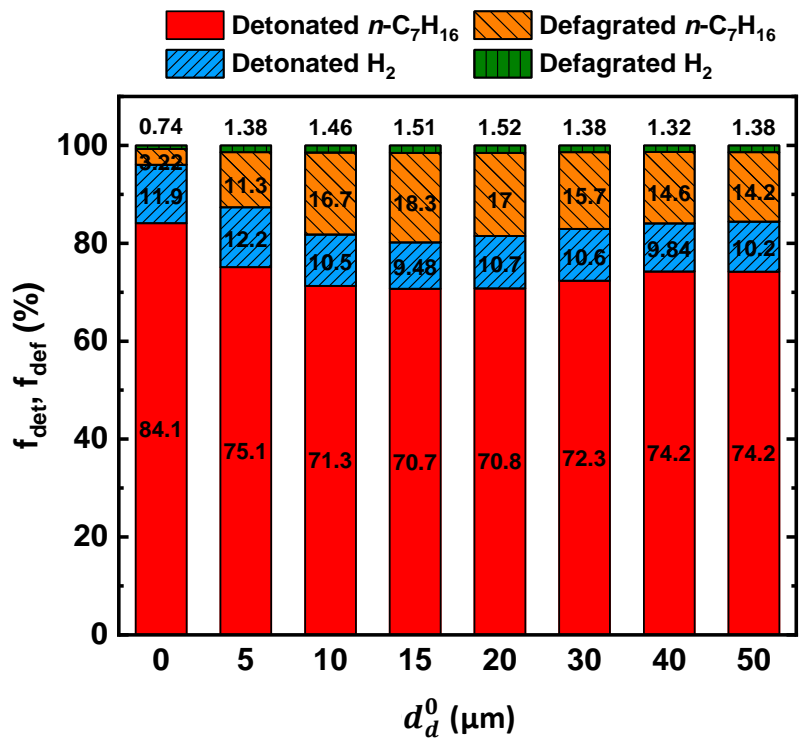

Fig. 10 Detonated and deflagrated fuel fractions. Results from case group B ($\phi_g = 0.4$).

The fractions of detonated ($f_{det}$) and deflagrated ($f_{def}$) fuels are further examined, based on the cases of $\phi_g$ = 0.4 (group B). The detonated (deflagrated) fuel fraction is the fuel consumption rate of detonative (deflagrative) combustion, normalized by the sum of the overall fuel consumption rates, i.e. $\dot{\omega}_{n-C7H16} + \dot{\omega}_{H2}$. Here the deflagration is identified based on $\dot{Q} \leq 10^{13}$ J/m$^3$/s. It is seen that over 70% n-heptane is detonated under different droplet diameter conditions. The detonated n-heptane fraction first decreases until $d_d^0$ = 15 µm and then slightly increases. The fractions of deflagrated n-heptane have similar tendency but are much lower (less than 18%). Also, the detonated and deflagrated hydrogen fractions show limited variations when the droplet diameter is increased.

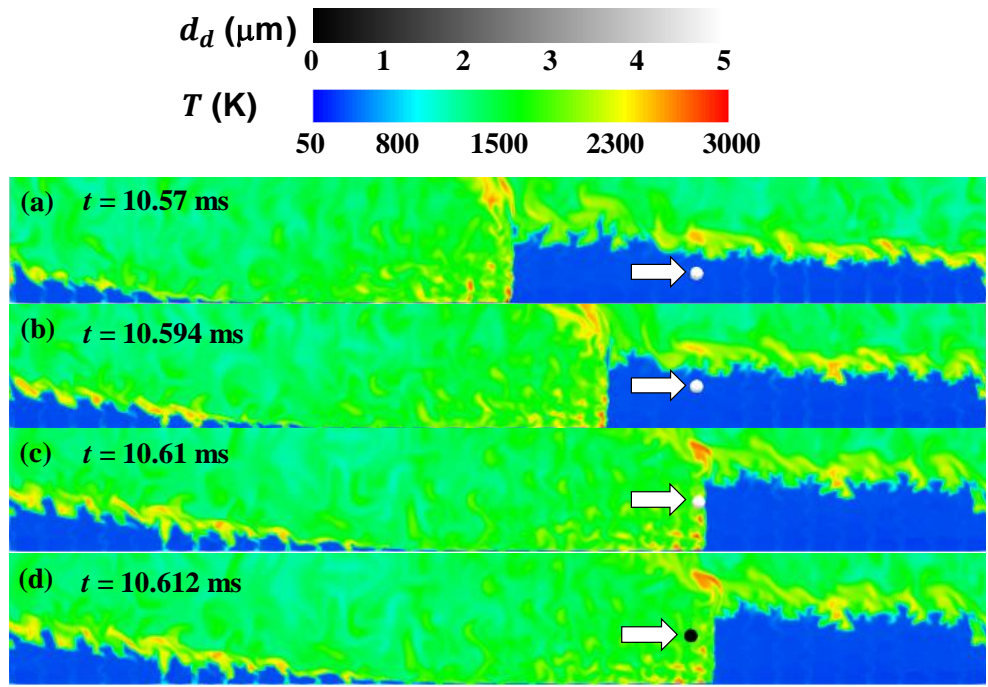

Fig. 11 Droplet trajectory I: be fully vaporized around the detonation front. Results from case B05. Arrows indicate the droplet locations. Image size: 288 mm × 40 mm.

## 5.4 Typical droplet trajectory in rotating detonative combustion

Four representative droplet trajectories in rotating detonation combustion are identified through screening the fates of a large number of droplets after they are injected into the domain. The first trajectory ("trajectory I" for reference hereafter) is shown in Figs. 11(a)−11(d), which show the time sequence of the instantaneous location of one liquid fuel droplet in case B05 ($\phi_g = 0.4$ and $d_d^0 = 5$ μm), and the gas temperature are overlaid to demonstrate the detonation wave location. Meanwhile, for this droplet, its time histories of diameter squared ($d_d^2$), evaporation rate ($\dot{m}_d$), $x$- and $y$-direction velocities ($U_x$ and $U_y$), magnitude of gas velocity ($U_c$) at the droplet location, and interphase velocity magnitude difference ($U_c - U_d$) are demonstrated in Fig. 12.

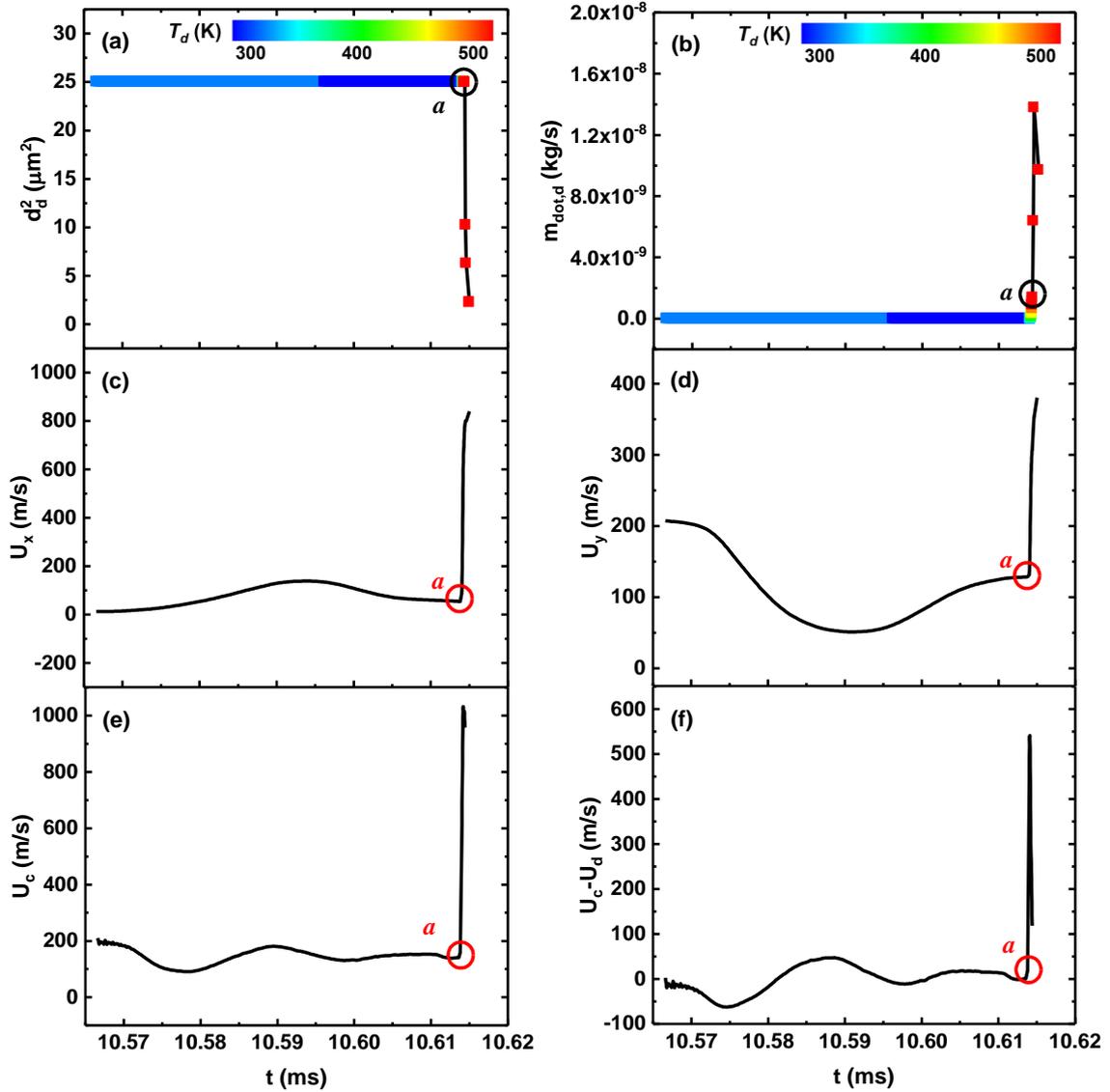

Fig. 12 Time histories of (a) diameter squared, (b) evaporation rate, (c) $x$-direction velocity, (d) $y$-direction velocity, (e) local gas velocity magnitude, (f) velocity difference in trajectory I.

It is seen from Fig. 11(a) that this droplet moves mainly along the RDE height direction in the refill zone after it is injected at $t$ = 10.567 ms. This can be confirmed by relatively small $x$-direction and dominant $y$-direction velocities, as shown in Fig. 12(c) and 12(d) respectively. The $y$-direction velocity is around 200 m/s at the inlet, decreases due to the drag effects and then increases towards 120 m/s before the detonation arrives. The increased velocity results from the recirculating flows near the walls, which can be observed in Fig. 5. The interphase velocity difference is small before the detonation front as shown in Fig. 12(f), implying the interphase kinematic quasi-equilibrium in the refill zone. Then the droplet stagnates some distance below the deflagrative surface, as seen in Fig. 11(b). During this period, its diameter $d_d$ almost keeps constant (see Fig. 12a) and the droplet temperature is relaxed to local gas temperature (about 250 K) due to the convective heat transfer. Limited evaporation happens and therefore almost zero $\dot{m}_d$ is observed in Fig. 12(b). At $t$ = 10.61 ms (marked with $a$ in Fig. 12), the droplet is swept by the approaching detonation wave (see Fig. 11c), and therefore rapidly heated to the boiling temperature (about 540 K, see the colouring of Figs. 12a and 12b). Strong evaporation proceeds (with high $\dot{m}_d$ in Fig. 12b) and the droplet is fully vaporized (with quickly decaying size in Fig. 12a) when it crosses the detonation wave.

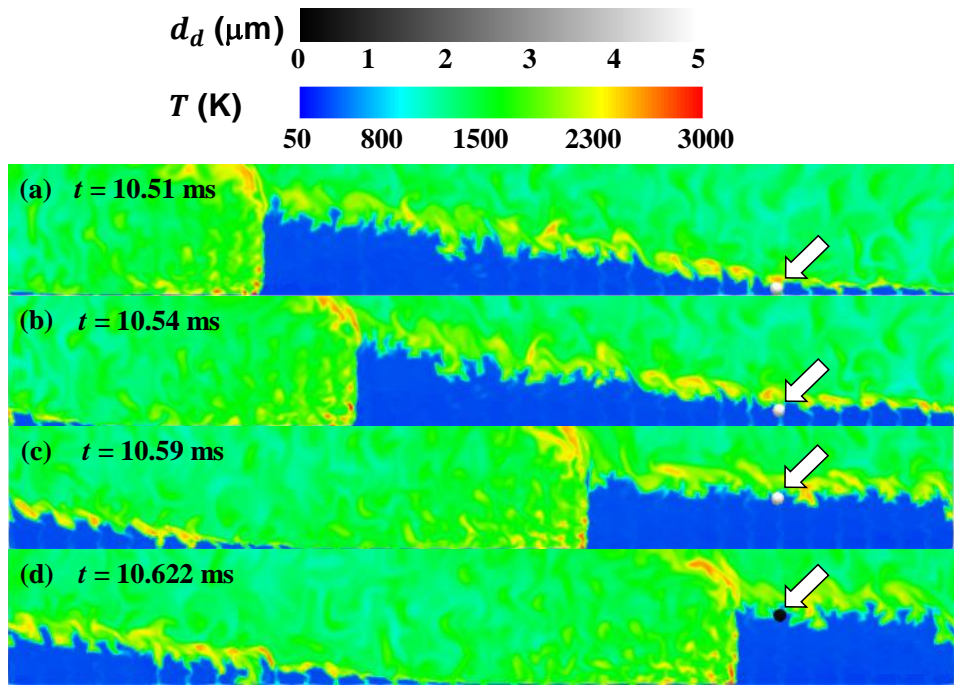

Fig. 13 Droplet trajectory II: be fully vaporized around the deflagrative surface. Results from case B05. Arrows indicate the droplet locations. Image size: 288 mm × 40 mm.

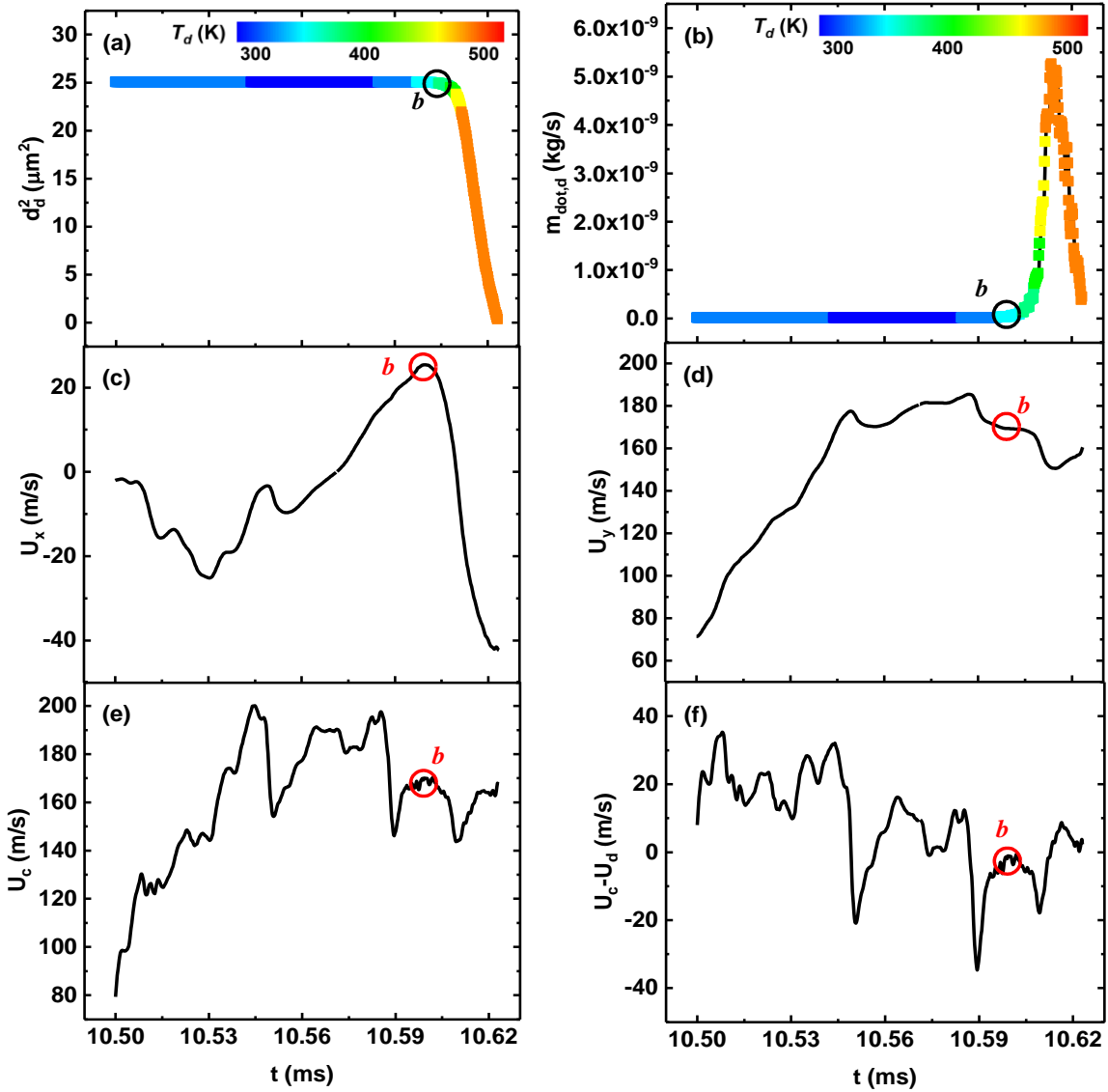

Fig. 14 Time histories of (a) diameter squared, (b) evaporation rate, (c) *x*-direction velocity, (d) *y*-direction velocity, (e) local gas velocity magnitude, (f) interphase velocity difference in trajectory II.

Likewise, Figs. 13 and 14 show the counterpart results of droplet trajectory II in rotating detonation field. The results are from the same case B05 as shown in Figs. 11 and 12. Different from trajectory I, this droplet is injected at $t = 10.5$ ms, from an inlet at the right end of the fuel refill zone and hence relatively far away from detonation wave front. Before $t =$ 10.605 ms (marked as *b* in Fig. 14), the droplet temperature is gradually reduced from the initial value of 300 K towards the local gas temperature (i.e. about 250 K) in the fill zone and then slightly increases when it approaches the hot deflagration surface. This trend is also seen from trajectory I in Fig. 12. Meanwhile, the droplet size and evaporation rate do not show pronounced changes (see Figs. 14a and 14b) in this period. However, considerable variations of droplet and gas velocities, as well as their difference, can be observed in Figs. 14(c)−14(f). These are different from the results in Fig 12. This is caused by its interactions with the eddies arising from the local deflagrative front instabilities. After $t =$ 10.605 ms, droplet temperature and evaporation rate $\dot{m}_d$ increase quickly (see Figs. 14b). This stems from the close interactions between the droplet and hot deflagrative surface, as indicated in Figs. 13(c) and 13(d). However, the droplet temperature stays around 500 K. After $t =$ 10.605 ms, droplet evaporation rate decreases because of the reduced droplet size, as demonstrated in Figs. 14(a) and 14(b).

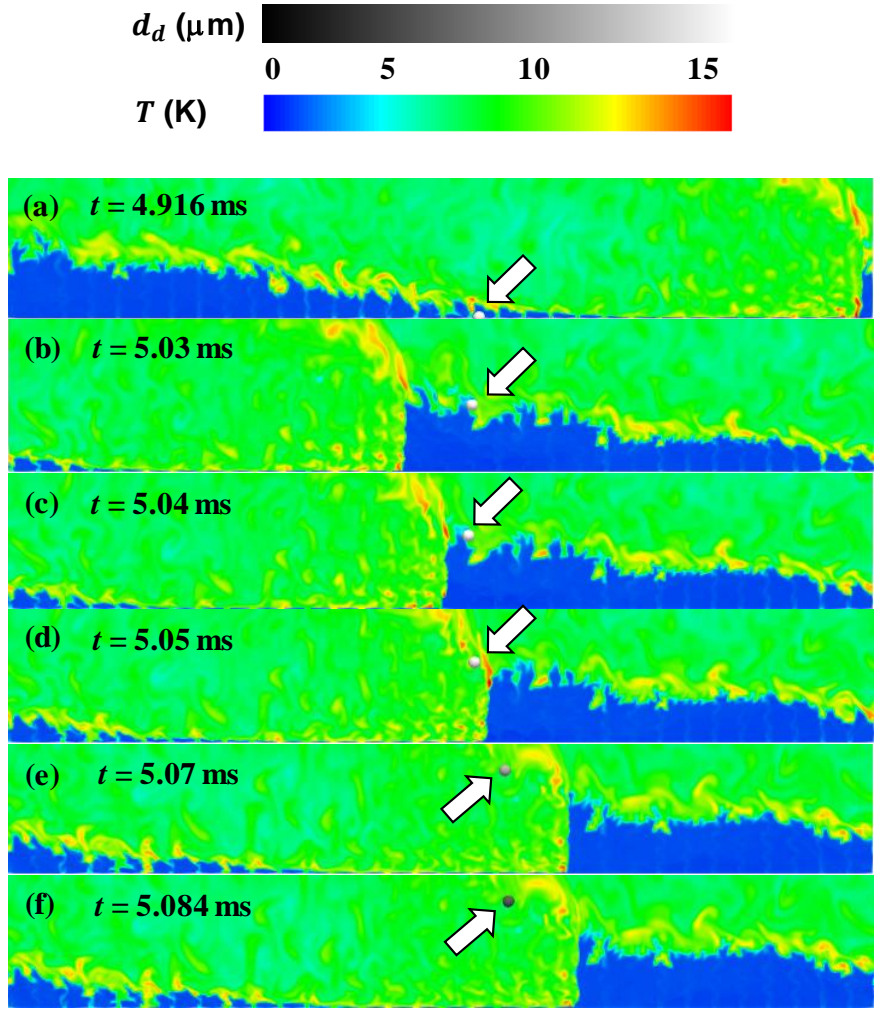

Fig. 15 Droplet trajectory III: leak through the triple point. Results from Case B15. Arrows indicate the droplet location. Image size: 288 mm × 40 mm.

Different from droplet trajectories I and II, Figs. 15 and 16 show the time evolutions of trajectory III, in which the droplet first interacts with the deflagrative surface and then leaves the refill zone through the triple point. The results are from case B15 and the initial droplet diameter is $d_d^0 = 15$ μm, larger than those in trajectories I and II. Similar to the droplet in trajectory II, this droplet is also injected far from the propagating detonation wave. At its early stage ($t \leq 5.03$ ms), the droplet temperature decreases gradually towards local gas temperature, and the evaporation rate $\dot{m}_d$ is almost zero (see Fig. 16b), as observed in trajectories I and II. Nevertheless, when the droplet approaches the triple point (e.g. at $t = 5.04$ ms, marked as $c$ in Fig. 16), its temperature, velocity and evaporation rate abruptly increase, as seen from Fig. 16. Considerable interphase velocity difference is observed from Fig. 16(f) at that instant. When it enters the burned area, like at $t$ = 5.05 ms and 5.07 ms in Figs. 15(d) and 15(e), the evaporation is greatly enhanced, leading to considerably reduced droplet size. This can be seen from Figs. 16(a) and 16(b). The oscillation of $\dot{m}_d$ results from the spatially non-uniform thermo-chemical compositions (e.g. temperature) behind the detonation wave. Meanwhile, the droplet velocities are generally high, in line with the local gas flows near the slip line (hence low interphase velocity difference). Finally, the droplet is completely vaporized at $t$ = 5.084 ms in Fig. 15(f).

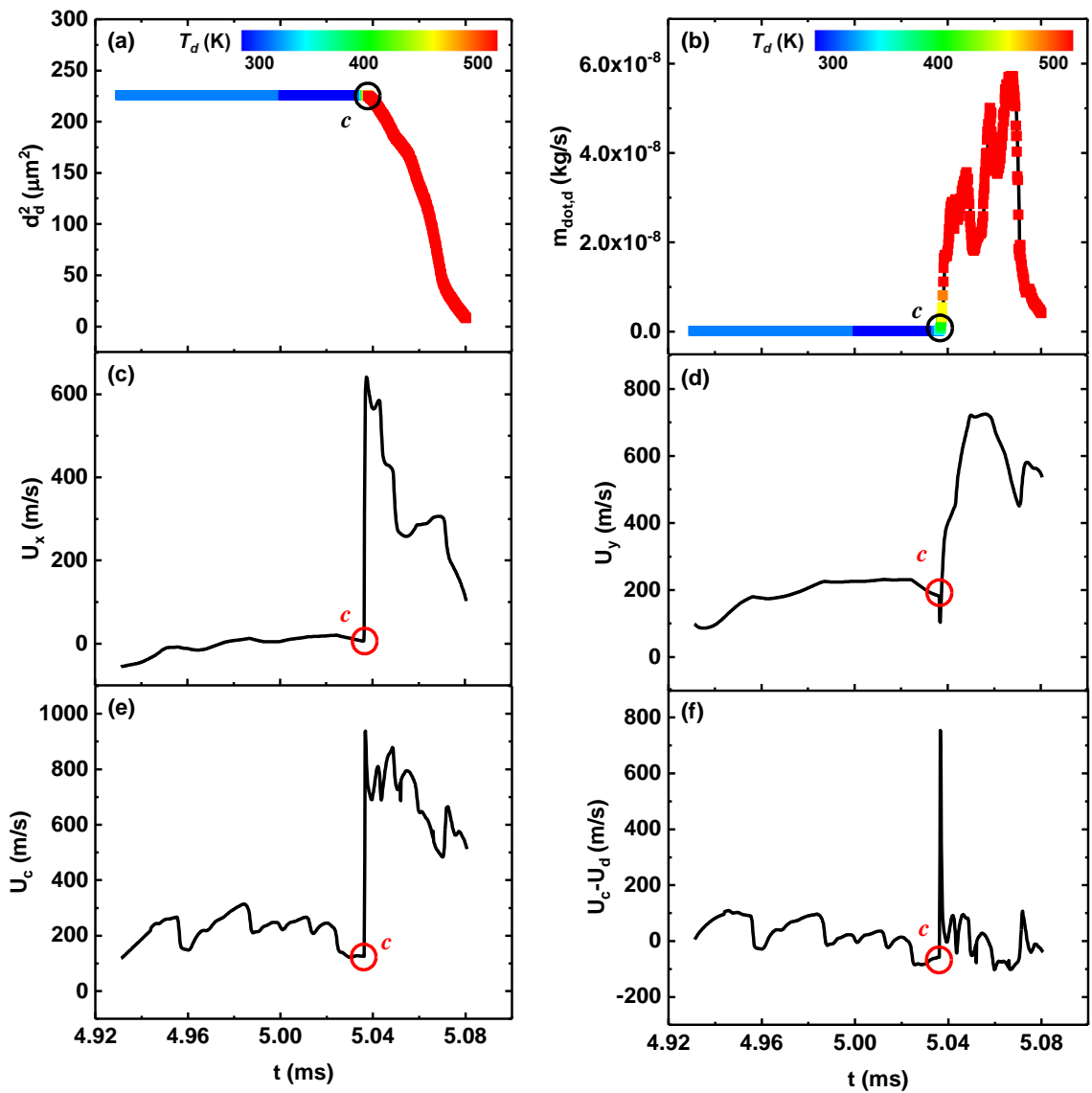

Fig. 16 Time histories of (a) diameter squared, (b) evaporation rate, (c) *x*-direction velocity, (d) *y*-direction velocity, (e) local gas velocity magnitude, (f) interphase velocity difference in trajectory III.

The last type of droplet trajectory observed from our results is shown in Figs. 17 and 18, termed as droplet trajectory IV. The results are from the three-waved case C40 with $\phi_g = 0.6$ and $d_0 = 40$ μm. After the droplet is injected from inlet at $t = 11.58$ ms, it travels through the detonation front immediately (see Figs. 17a and 17b), which marked as *d* in Fig. 18. In the detonated gas, the droplet temperature quickly reaches the boiling point and evaporation rate increases accordingly, peaking at about $t = 11.60$ ms. This can be clearly seen from Figs. 18(a) and 18(b). Between $t = 11.59$ ms and 11.65 ms, the droplet diameter and *x*-direction velocity continuously decrease. However, the *y*-direction velocity increases due to the expansion of the detonation product gas. The second milestone of this droplet is to cross the oblique shock wave connecting with the next detonation wave, and this instant is marked as *e* in Fig. 18. This is manifested by abruptly increased droplet and gas velocities, evaporation rate $\dot{m}_d$ also slightly increases, due to the high temperature in the post-shock zone, but decays quickly due to the quickly reduced droplet diameter.

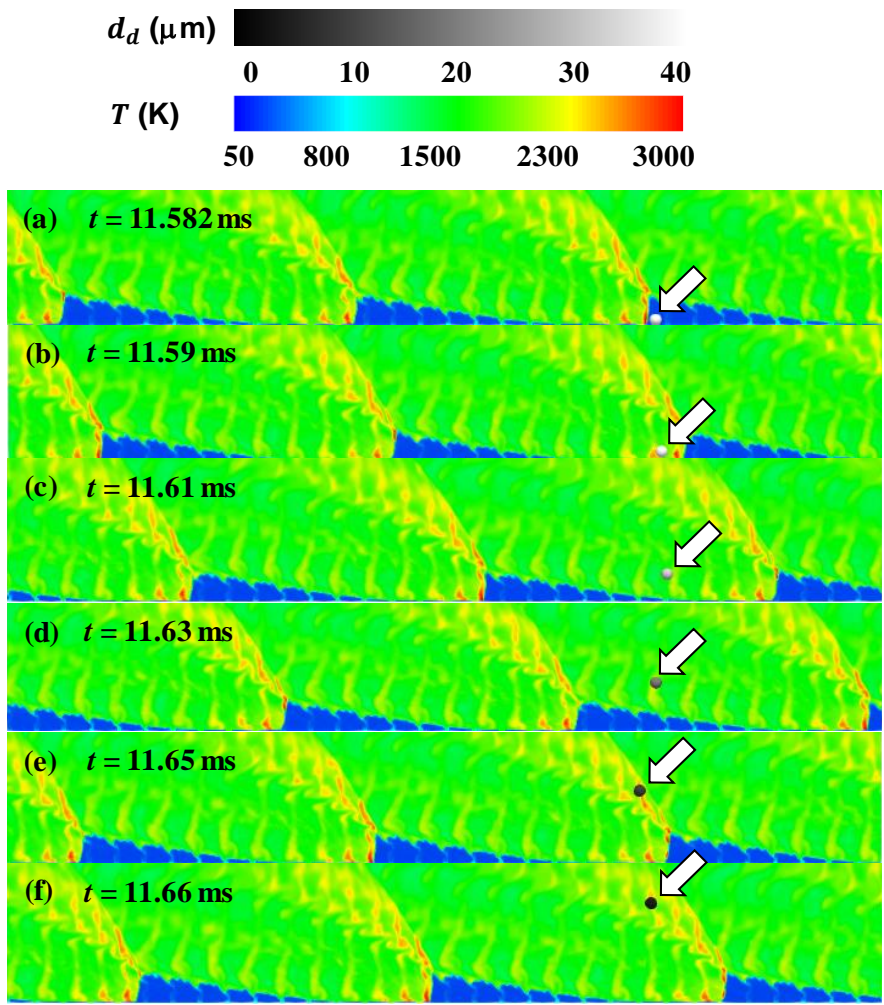

Fig. 17 Droplet trajectory IV: cross the detonation wave and interact with the shock wave. Results from case C40. Arrows indicate the droplet location. Image size: 288 mm × 40 mm.

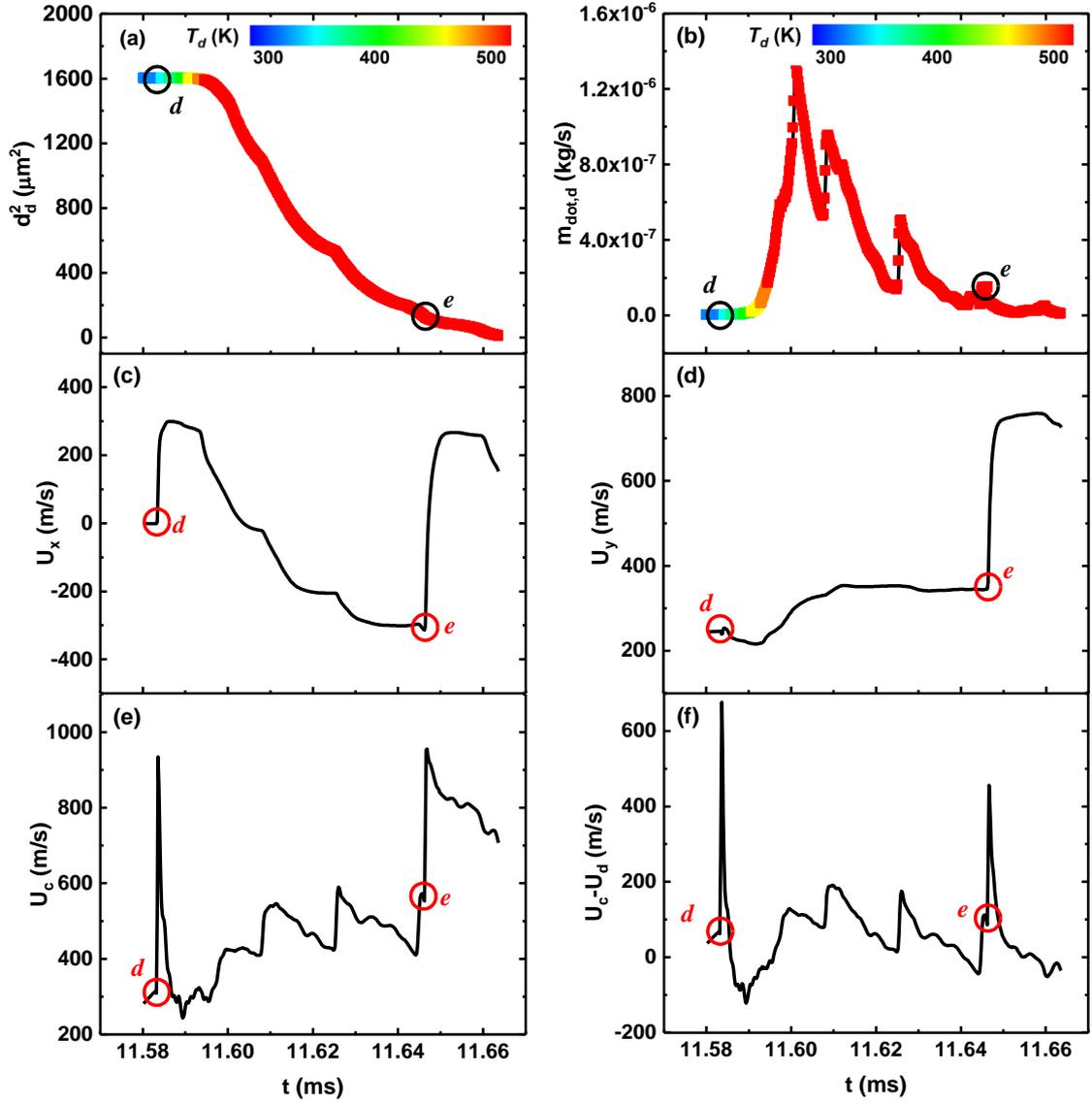

Fig. 18 Time histories of (a) diameter squared, (b) evaporation rate, (c) *x*-direction velocity, (d) *y*-direction velocity, (e) local gas velocity magnitude, (f) interphase velocity difference in trajectory IV.

Table 4 summarizes the key time and length scales of droplet and rotating detonation wave in the foregoing four trajectories. Several interesting phenomena merit further discussion here. Firstly, the computed evaporation time of trajectory I is $\tau_{ev}^{sim} \approx 2$ μs, much shorter than the theoretical value, i.e. $\tau_{ev}^{est} \approx 5$ μs, calculated from droplet evaporation in stagnant flows and therefore $Sh \approx 2.0$ [62]

$$\tau_{ev}^{est} = \frac{\rho_d {d_d^0}^2}{8\rho_s D_{ab} ln(1 + X_r)}, \qquad (35)$$

where $\rho_s$, $X_r$ and $D_{ab}$ are estimated with the droplet evaporation temperature. Moreover, in trajectory II, $\tau_{ev}^{sim} = 20$ μs is close to $\tau_{ev}^{est}$ from Eq. (35). This is justifiable since the droplet has relatively low interphase velocity difference (low $Re_d$ and therefore a good approximation of $Sh \approx 2$). However, in trajectories III and IV, $\tau_{ev}^{sim}$ are 33% and 13% higher than $\tau_{ev}^{est}$, respectively. Different from the droplets in trajectories I and II, those in trajectories III and IV mainly travel beyond the refill zone. This may be affected by the variable surrounding gas atmosphere behind the detonation wave as illustrated in Figs. 15 and 17. However, Eq. (35) still provides good estimates for droplet evaporation time in rotation detonation combustion.

Table 4 Time and length scales in various droplet trajectories

| Trajectory | Droplet evaporation time (μs) | | Droplet residence time (μs) | | RDW speed $D_{det}$ (m/s) | Refill zone length $l_f$ (m) | Injection distance $l_i$ (m) |
|---|---|---|---|---|---|---|---|
| | Present simulation $\tau_{ev}^{sim}$ | Theoretical Estimation $\tau_{ev}^{est}$ | Whole domain $\tau_{res}^{tot}$ | Refill zone $\tau_{res}^{ref}$ | | | |
| I (5 μm) | 2 | 5 | 48 | 46 | 1,390 | 0.193 | 0.064 |
| II (5 μm) | 20 | 22 | 123 | 123 | 1,390 | 0.193 | 0.193 |
| III (15 μm) | 40 | 30 | 149 | 107 | 1,376 | 0.200 | 0.158 |
| IV (40 μm) | 76 | 67 | 82 | 4 | 1,516 | 0.060 | 0.004 |

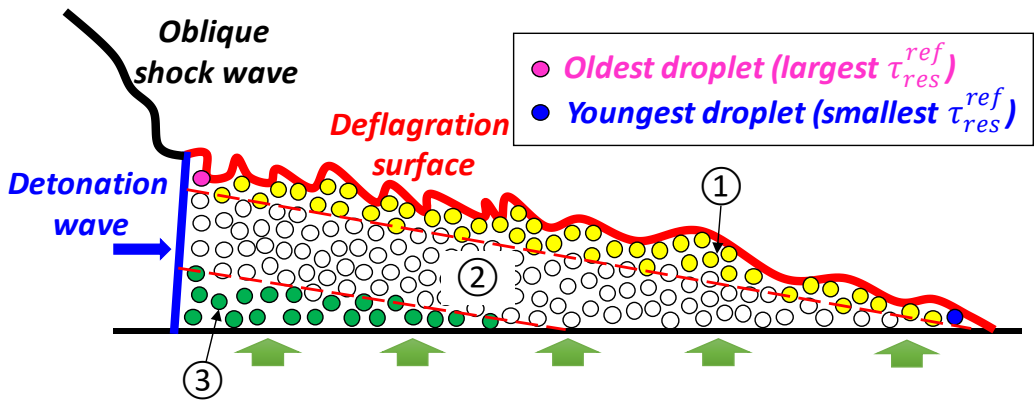

Fig. 19 Schematic of droplet distribution in the fuel refill zone of rotating detonation combustion.

Secondly, it is of great relevance to enhance the droplet residence time in the fuel refill zone in order to increase the liquid fuel utilization and detonated fuel fraction in RDE's. As listed in Table 4, the total residence times $\tau_{res}^{tot}$ of droplets in the four trajectories are 48, 123, 149 and 82 μs respectively, and the refill zone residence time $\tau_{res}^{ref}$ respectively accounts for about 96%, 100%, 72% and 5% of them. Note that if $\tau_{res}^{ref} < \tau_{res}^{tot}$ (the droplet does not fully vaporize in the refill zone), then $\tau_{res}^{ref}$ can be approximated from $\tau_{res}$, i.e.

$$\tau_{res}^{ref} \approx \tau_{res} = \frac{l_i}{D_{det}}, \tag{36}$$

where $l_i$ is the distance between the droplet injector and the instantaneous detonation wave and $l_i$ is always less than the refill zone length, i.e. $l_i \leq l_f$. Equation (36) holds when the *y*-direction velocity component is low and its validity can be confirmed from close values of $\tau_{res}$ and $\tau_{res}^{ref}$ in Table 4. Figure 19 schematically shows the distribution of liquid fuel droplets in the triangular refill zone, and three categories of droplets can be loosely grouped, marked with 1, 2 and 3. The first category of droplets (in yellow in Fig. 19) generally have high $\tau_{res}^{ref}$, which lies at the tips of the spray jets and are injected near the right end of the instantaneous refill zone (hence with large $l_i$). Two extremes, the oldest and youngest droplets (in pink and blue respectively), are near the triple point and at the rightmost end of the zone, respectively. The duration with which they interact with, or affected by, the deflagration surface can be approximated as $\tau_{res}^{ref}$. These droplets may be fully vaporized inside the refill zone (e.g. trajectory II), or escape near the triple point (e.g. trajectory III), and have significant influence on detonated fuel fraction as discussed in Section 5.3. Either fate is expected to have small kinetic contributions for detonative combustion, and therefore should be avoided in practical liquid RDE technologies.

For the third droplet category (in green), they are just injected into the refill zone and featured by small $l_i$ and $\tau_{res}^{ref}$ (e.g. trajectory IV). Accordingly, unless they are perfectly atomized and sprayed with small $d_d^0$, they would have limited time to be heated and vaporized, leading to negligible vapour addition and subsequent mixing with oxidant for the encroaching RDW. On the contrary, the evaporative cooling and interphase momentum exchange may adversely affect the stable propagation of detonation wave base near the injector. The second category of droplets with intermediate $l_i$ have longer $\tau_{res}^{ref}$ (compared to the second category) and do not directly interact with deflagration surface (compared to the first category). It is expected to play a major role in providing vaporized fuel before or on RDW arrival. Therefore, how to implement and optimize *in-situ* droplet evaporation inside the refill zone of practical RDE combustors deserves further investigations using both computational and experimental methods.

*5.5 Droplet dispersion height in rotating detonation combustor*

Figure 20 shows the variations of Sauter Mean Diameter (SMD, $d_{32}$) along the RDE height direction (*y*-direction) for different pre-vaporized gas equivalence ratios and initial droplet diameters. Here SMD is calculated based on the ratio of the volume to surface area of the droplets distributed along the entire *x*-direction at a fixed height. Note that the mean detonation wave height is marked by a rectangle and the entire RDE height is *y* = 99 mm. It is seen that the droplet SMD are almost unchanged initially when the axial distance is less than the RDW height. This is because most of these droplets are within the fuel refill zone and limited droplet vaporization occurs due to the low gas temperature. The SMD decreases rapidly when the droplets are beyond the detonation wave height. This can be justified by the fact that the droplets are surrounded by hot detonation product and the droplets keep evaporating there. Small-sized droplets are completely vaporized before arriving the outlet (*y* = 99 mm). However, larger droplets escape from outlet and this would weaken the fuel utilization. It is also found that all the droplets are vaporized within the RDE combustor for high $\phi_g$ is 0.6 and above (see Figs. 20c−20e). Therefore, increasing droplet pre-evaporation in the liquid fuel supply can promote the *in-situ* vaporization of the liquid droplets inside the chamber.

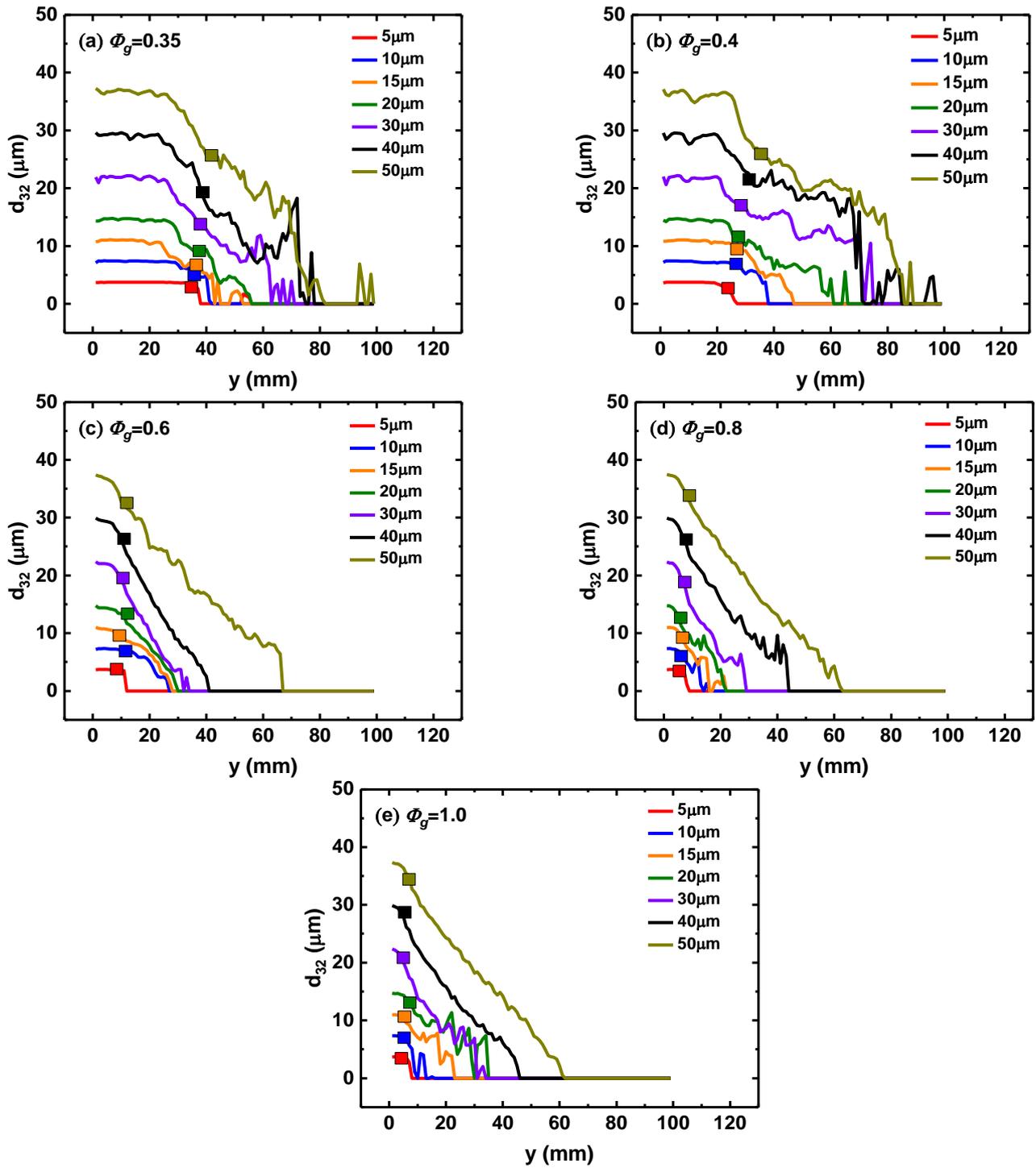

Fig. 20 Profiles of Sauter mean diameter along the RDE height direction.

Figure 21 further shows the variations of the droplet dispersion height ($H$) for all the simulated cases. $H$ is defined as the $y$-direction distance between the RDE head end and the critical axial height beyond which no droplets are dispersed (i.e. $d_{32} = 0$ in Fig. 20). We can find that $H$ increases monotonically as the initial droplet size $d_d^0$ increases for a fixed pre-vaporized equivalence ratio $\phi_g$ (i.e. within the same group). In addition, with increased $\phi_g$ from groups A to E, $H$ decreases gradually. If the droplets cannot be completely vaporized before the RDE exit, then we assume $H$ = 99 mm, which may occur when $d_d^0$ is large and/or $\phi_g$ is small, e.g. cases A50, A40, B50 and B40 as shown in Fig. 21. This does not only deteriorate the fuel utilization but also adversely affects or even damages the downstream structures, e.g. exit nozzle or turbine blades when the detonation combustor is integrated with turbomachinery. Therefore, from RDE design perspective, besides the general requirements of the detonation wave height discussed by Bykovskii et al. [58], optimal chamber height of liquid fuelled RDE should be determined, by considering the fuel atomization and evaporation characteristics. This topic merits further studies in the future work.

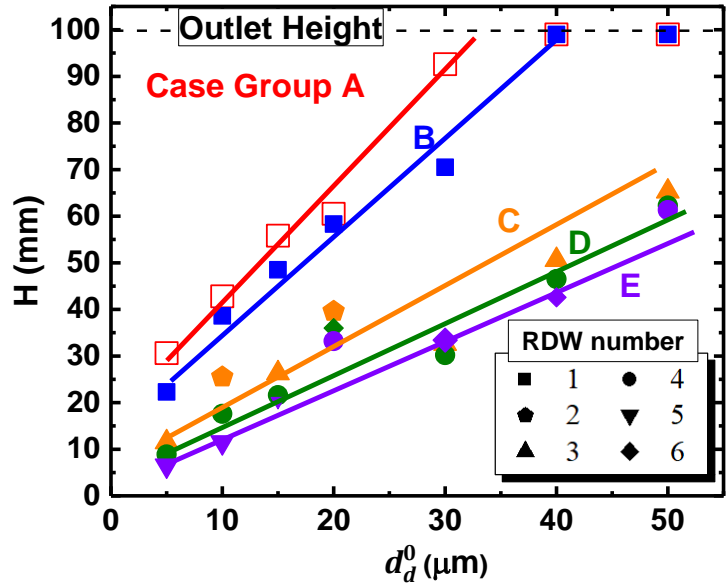

Fig. 21 Change of droplet dispersion height with initial droplet diameter.

## 6. Conclusions

Eulerian－Lagrangian modelling of two-dimensional rotating detonative combustion fuelled with partially pre-vaporized *n*-heptane sprays and gaseous hydrogen is performed. Our emphasis is laid on detonation wave propagation in heterogeneous mixtures and droplet dynamics in rotating detonation combustion. Different pre-vaporized *n*-heptane equivalence ratios ($\phi_g$) and droplet diameters ($d_d^0$) are investigated.

The results show that when the droplets are small, they are fully vaporized by the detonation wave. However, when the diameter is relatively large and/or the detonation wave number is bifurcated, liquid droplets are observable beyond the refill zone. They continue evaporating in the post-detonated gas and the vapour is deflagrated. Moreover, the detonation propagation speed is increased with $\phi_g$. When $\phi_g$ is low (e.g. 0.35 and 0.4), the propagation speed is enhanced in small-sized sprays, compared to the purely gaseous detonation. The velocity deficit increases with $d_d^0$ and/or $\phi_g$, and varies from 5% to 30% in our studied cases.

It is also found that the detonated fuel fraction increases with $\phi_g$. However, there exists a critical droplet diameter (about 20 µm), with which the detonated fuel fraction is lowest. Over 70% *n*-heptane is detonated in the simulated cases. Four typical droplet trajectories in rotating detonation combustion are identified, and they are characterized by various evaporation times, residence times and their interactions with the basic structures in the detonation combustion field. Inside the refill zone, three droplet categories are qualitatively identified. Droplets injected at the right end of the refill zone directly interact with the deflagration surface and have relatively long residence time. However, droplets injected closer to the detonation front have insufficient time to be heated and vaporized in the refill zone. Further studies are needed to develop efficient technologies to enhance the *in-situ* droplet evaporation during the refill period for practical RDE implementations.

Our results also demonstrate that when $\phi_g$ is low (0.35 and 0.4) and $d_d^0$ is large (40 and 50 µm), the liquid fuel droplets may disperse toward the domain exit. When $\phi_g$ is high, they can be fully evaporated inside the chamber. Furthermore, the droplet dispersion height is reduced when the pre-evaporation degree becomes high, whilst it increases when $d_d^0$ is increased.

**Acknowledgments**

This work used the computational resources of the National Supercomputing Centre, Singapore (https://www.nscc.sg). QM is supported by the China Scholarship Council (No. 201906680008).

# References

[1] V. Anand, E. Gutmark, Rotating detonation combustors and their similarities to rocket instabilities, Prog. Energy Combust. 73 (2019) 182-234.
[2] P. Wolański, Detonative propulsion, Proc. Combust. Inst. 34 (2013) 125-158.
[3] B.A. Rankin, D.R. Richardson, A.W. Caswell, A.G. Naples, J.L. Hoke, F.R. Schauer, Chemiluminescence imaging of an optically accessible non-premixed rotating detonation engine, Combust. Flame 176 (2017) 12-22.
[4] A. Kawasaki, T. Inakawa, J. Kasahara, K. Goto, K. Matsuoka, A. Matsuo, I. Funaki, Critical condition of inner cylinder radius for sustaining rotating detonation waves in rotating detonation engine thruster, Proc. Combust. Inst. 37 (2019) 3461-3469.
[5] R. Bluemner, M.D. Bohon, C.O. Paschereit, E.J. Gutmark, Effect of inlet and outlet boundary conditions on rotating detonation combustion, Combust. Flame 216 (2020) 300-315.
[6] Y. Zheng, C. Wang, Y. Wang, Y. Liu, Z. Yan, Numerical research of rotating detonation initiation processes with different injection patterns, Int. J. Hydrogen Energy 44 (2019) 15536-15552.
[7] R. Driscoll, P. Aghasi, A. St George, E.J. Gutmark, Three-dimensional, numerical investigation of reactant injection variation in a h2/air rotating detonation engine, Int. J. Hydrogen Energy 41 (2016) 5162-5175.
[8] D.A. Schwer, K. Kailasanath, Towards non-premixed injection modeling of rotating detonation engines, 51$^{st}$ AIAA/SAE/ASEE Joint Propulsion Conference (2015), paper 3782.
[9] B. Sun, H. Ma, Two-dimensional numerical study of two-phase rotating detonation wave with different injections, AIP Adv. 9 (2019) 115307.
[10] M. Zhao, J.M. Li, C.J. Teo, B.C. Khoo, H. Zhang, Effects of variable total pressures on instability and extinction of rotating detonation combustion, Flow Turbul. Combust. 104 (2020) 261-290.
[11] T.H. Yi, J. Lou, C. Turangan, J.Y. Choi, P. Wolanski, Propulsive performance of a continuously rotating detonation engine, J. Propuls. Power 27 (2011) 171-181.
[12] N. Jourdaine, N. Tsuboi, K. Ozawa, T. Kojima, A.K. Hayashi, Three-dimensional numerical thrust performance analysis of hydrogen fuel mixture rotating detonation engine with aerospike nozzle, Proc. Combust. Inst. 37 (2019) 3443-3451.
[13] V.R. Katta, K.Y. Cho, J.L. Hoke, J.R. Codoni, F.R. Schauer, W.M. Roquemore, Effect of increasing channel width on the structure of rotating detonation wave, Proc. Combust. Inst. 37 (2019) 3575-3583.
[14] T. Kaemming, M.L. Fotia, J. Hoke, F. Schauer, Thermodynamic modeling of a rotating detonation engine through a reduced-order approach, J. Propuls. Power 33 (2017) 1170-1178.
[15] C.A. Nordeen, D. Schwer, F. Schauer, J. Hoke, T. Barber, B. Cetegen, Thermodynamic model of a rotating detonation engine, Combust. Explos. Shock 50 (2014) 568-577.
[16] P.W. Shen, T. Adamson Jr, Theoretical analysis of a rotating two-phase detonation in liquid rocket motors, Astronaut. Acta 17 (1972) 715-728.
[17] R. Zhou, J.P. Wang, Numerical investigation of flow particle paths and thermodynamic performance of continuously rotating detonation engines, Combust. Flame 159 (2012) 3632-3645.
[18] F.A. Bykovskii, S.A. Zhdan, E.F. Vedernikov, Continuous spin detonation of fuel-air mixtures, Explos. Shock Waves 42 (2006) 463-471.
[19] F.A. Bykovskii, S.A. Zhdan, E.F. Vedernikov, Continuous detonation of the liquid kerosene—air mixture with addition of hydrogen or syngas, Combust. Explos. Shock 55 (2019) 589-598.
[20] F.A. Bykovskii, S.A. Zhdan, E.F. Vedernikov, Continuous spin detonation of a heterogeneous kerosene–air mixture with addition of hydrogen, Combust. Explos. Shock 52 (2016) 371-373.
[21] J. Kindracki, Experimental studies of kerosene injection into a model of a detonation chamber, J. Power Technol. 92 (2012) 80-89.
[22] J. Kindracki, Experimental research on rotating detonation in liquid fuel–gaseous air mixtures, Aerosp. Sci. Technol. 43 (2015) 445-453.
[23] J.M. Li, P.H. Chang, L. Li, Y. Yang, C.J. Teo, B.C. Khoo. Investigation of injection strategy for liquid-fuel rotating detonation engine, AIAA Aerospace Sciences Meeting (2018), paper 0403.
[24] S. Xue, H. Liu, L. Zhou, W. Yang, H. Hu, Y. Yan, Experimental research on rotating detonation with liquid hypergolic propellants, Chin. J. Aeronaut. 31 (2018) 2199-2205.
[25] A.K. Hayashi, N. Tsuboi, E. Dzieminska, Numerical study on JP-10/air detonation and rotating detonation engine, AIAA J. (2020) 1-17.

[26] M. Zhao, H. Zhang, Modelling rotating detonative combustion fueled by partially pre-vaporized n-heptane sprays, Accepted by 38[th] International Symposium on Combustion, under reviewed by Proc. Combust. Inst. (2021).

[27] W. Sutherland, LII. The viscosity of gases and molecular force, The London, Edinburgh, and Dublin Philosophical Magazine and Journal of Science 36 (2009) 507-531.

[28] R.C. Reid, J.M. Prausnitz, B.E. Poling, The properties of gases and liquids, McGraw-Hill, New York, U.S.A., 2007.

[29] B.J. McBride, Coefficients for calculating thermodynamic and transport properties of individual species, Report No. NASA-TM-4513, NASA Lewis Research Center, Cleveland, OH, USA, 1993.

[30] G.B. Macpherson, N. Nordin, H.G. Weller, Particle tracking in unstructured, arbitrary polyhedral meshes for use in cfd and molecular dynamics, Commu. Numer. Mechods Eng. 25 (2009) 263-273.

[31] C.T. Crowe, J.D. Schwarzkopf, M. Sommerfeld, Y. Tsuji, Multiphase flows with droplets and particles, CRC Press, Boca Raton, USA, 2011.

[32] A.B. Liu, D. Mather, R.D. Reitz, Modeling the effects of drop drag and breakup on fuel sprays, 102 (1993) 83-95.

[33] M. Doble, Perry's chemical engineers' handbook, McGraw-Hill, New York, U.S.A., 2007.

[34] W. Ranz, W.R. Marshall, Evaporation from drops, 48 (1952) 141-146.

[35] H.G. Weller, G. Tabor, H. Jasak, C. Fureby, A tensorial approach to computational continuum mechanics using object-oriented techniques, 12 (1998) 620-631.

[36] C.J. Greenshields, H.G. Weller, L. Gasparini, J.M. Reese, Implementation of semi-discrete, non-staggered central schemes in a colocated, polyhedral, finite volume framework, for high-speed viscous flows, Int. J. Numer. Methods Fluids, 63 (2010) 1-21.

[37] A. Kurganov, S. Noelle, G. Petrova, Semidiscrete central-upwind schemes for hyperbolic conservation laws and hamilton--jacobi equations, 23 (2001) 707-740.

[38] M. Zhao, H. Zhang, Origin and chaotic propagation of multiple rotating detonation waves in hydrogen/air mixtures, Fuel 275 (2020) 117986.

[39] L.G. Marcantoni, J. Tamagno, S. Elaskar, Rhocentralrffoam: An openfoam solver for high speed chemically active flows–simulation of planar detonations–, 219 (2017) 209-222.

[40] Z. Huang, M.Zhao, Y. Xu, G. Li, H. Zhang, Eulerian-Lagrangian modelling of detonative combustion in two-phase gas-droplet mixtures with OpenFOAM: Validations and verifications, Fuel 286 (2021) 119402.

[41] F. Ma, J.Y. Choi, V. Yang, Propulsive performance of airbreathing pulse detonation engines, J. Propuls. Power 22 (2006) 1188-1203.

[42] C.K. Westbrook, F.L. Dryer, Simplified reaction mechanisms for the oxidation of hydrocarbon fuels in flames, Combust. Sci. Technol. 27 (2007) 31-43.

[43] Meng Q, Zhao N, Zheng H, Yang J, Li Z, Deng F, A numerical study of rotating detonation wave with different numbers of fuel holes, Aerosp. Sci. Technol. 93 (2019) 105301.

[44] S. Liu, J.C. Hewson, J.H. Chen, H. Pitsch, Effects of strain rate on high-pressure nonpremixed n-heptane autoignition in counterflow, Combust. Flame 137 (2004) 320-339.

[45] D.A. Schwer, Multi-dimensional simulations of liquid-fueled JP10/oxygen detonations, AIAA Propulsion and Energy 2019 Forum (2019), paper 4042.

[46] Z. Ren, B. Wang, G. Xiang, L. Zheng, Effect of the multiphase composition in a premixed fuel–air stream on wedge-induced oblique detonation stabilisation, J. Fluid Mech. 846 (2018) 411-427.

[47] M.D. Bohon, R. Bluemner, C.O. Paschereit, E.J. Gutmark, High-speed imaging of wave modes in an rdc, Exp. Therm. Fluid Sci. 102 (2019) 28-37.

[48] L. Deng, H. Ma, C. Xu, X. Liu, C. Zhou, The feasibility of mode control in rotating detonation engine, Appl. Therm. Eng. 129 (2018) 1538-1550.

[49] L. Deng, H. Ma, C. Xu, C. Zhou, X. Liu, Investigation on the propagation process of rotating detonation wave, Acta Astronaut. 139 (2017) 278-287.

[50] S. Yao, J. Wang, Multiple ignitions and the stability of rotating detonation waves, Appl. Therm. Eng. 108 (2016) 927-936.

[51] S. Yao, M. Liu, J. Wang, Numerical investigation of spontaneous formation of multiple detonation wave fronts in rotating detonation engine, Combust. Sci. Technol. 187 (2015) 1867-1878.

[52] D. Schwer, K. Kailasanath, Numerical investigation of the physics of rotating-detonation-engines, Proc. Combust. Inst. 33 (2011) 2195-2202.

[53] M. Hishida, T. Fujiwara, P. Wolanski, Fundamentals of rotating detonations, Shock Waves 19 (2009) 1-10.

[54] N. Tsuboi, Y. Watanabe, T. Kojima, A.K. Hayashi, Numerical estimation of the thrust performance on a rotating detonation engine for a hydrogen–oxygen mixture, Proc. Combust. Inst. 35 (2015) 2005-2013.

[55] J. Shepherd, Copyright © 1993-2020 by California Institute of Technology, https://shepherd.Caltech.Edu/edl/publicresources/sdt/.

[56] H. Watanabe, A. Matsuo, K. Matsuoka, A. Kawasaki, J. Kasahara, Numerical investigation on propagation behavior of gaseous detonation in water spray, Proc. Combust. Inst. 37 (2019) 3617-3626.

[57] D.A. Schwer, E. O'Fallon Jr, D. Kessler. Liquid-fueled detonation modeling at the us naval research laboratory, Report No. NRL/MR/6041--18-9816, Naval Research Laboratory, Washington, D.C., USA, 2018.

[58] F.A. Bykovskii, S.A. Zhdan, E.F. Vedernikov, Continuous spin detonations, J. Propuls. Power 22 (2006) 1204-1216.

[59] Q. Li, P. Liu, H. Zhang, Further investigations on the interface instability between fresh injections and burnt products in 2-d rotating detonation, Comput. Fluids 170 (2018) 261-272.

[60] K. Kailasanath, Liquid-fueled detonations in tubes, J. Propuls. Power 22 (2006) 1261-1268.

[61] S.A. Gubin, M. Sichel, Calculation of the detonation velocity of a mixture of liquid fuel droplets and a gaseous oxidizer, Combust. Sci. Technol. 17 (2007) 109-117.

[62] B. Rochette, E. Riber, B. Cuenot, Effect of non-zero relative velocity on the flame speed of two-phase laminar flames, Proc. Combust. Inst. 37 (2019) 3393-3400.